\documentclass[12pt,preprint]{aastex}

\begin{document}

\def\kms{km s$^{-1}$}
\def\msun{M$_{\odot}$}
\def\rsun{R$_{\odot}$}

\title{The ELM Survey. I.  A Complete Sample of Extremely Low Mass White Dwarfs\altaffilmark{*}}

\author{Warren R.\ Brown$^1$,
	Mukremin Kilic$^{1,}$\altaffilmark{**},
	Carlos Allende Prieto$^{2,3}$,
	and Scott J.\ Kenyon$^1$
	}

\affil{ $^1$Smithsonian Astrophysical Observatory, 60 Garden St, Cambridge, MA 02138\\
	$^2$Instituto de Astrof\'{\i}sica de Canarias, E-38205, La Laguna, Tenerife, Spain\\
	$^3$Departamento de Astrof\'{\i}sica, Universidad de La Laguna, E-38205 La Laguna, Tenerife, Spain
	}

\email{wbrown@cfa.harvard.edu, mkilic@cfa.harvard.edu, callende@iac.es, skenyon@cfa.harvard.edu}

\altaffiltext{*}{Based on observations obtained at the MMT Observatory, a joint 
facility of the Smithsonian Institution and the University of Arizona.}

\altaffiltext{**}{\em Spitzer Fellow}

\shorttitle{ Extremely Low Mass White Dwarfs }
\shortauthors{Brown et al.}

\begin{abstract}

	We analyze radial velocity observations of the 12 extremely low-mass
$\le$0.25 \msun\ white dwarfs (WDs) in the MMT Hypervelocity Star Survey.  Eleven of
the 12 WDs are binaries with orbital periods shorter than 14 hours; the one
non-variable WD is possibly a pole-on system among our non-kinematically selected
targets.  Our sample is unique:  it is complete in a well-defined range of apparent
magnitude and color.  The orbital mass functions imply that the unseen companions
are most likely other WDs, although neutron star companions cannot be excluded.  
Six of the 11 systems with orbital solutions will merge within a Hubble time due to
the loss of angular momentum through gravitational wave radiation.  The quickest
merger is J0923+3028, a $g=15.7$ ELM WD binary with a 1.08 hr orbital period and a
$\leq$130 Myr merger time. The chance of a supernova Ia event among our ELM WDs is
only 1\% -- 7\%, however.  Three binary systems (J0755+4906, J1233+1602, and
J2119$-$0018) have extreme mass ratios and will most likely form stable
mass-transfer AM CVn systems.  Two of these objects, SDSS J1233+1602 and
J2119$-$0018, are the lowest surface gravity WDs ever found; both show \hbox{Ca {\sc
ii}} absorption likely from accretion of circumbinary material.  We predict that at
least one of our WDs is an eclipsing detached double WD system, important for
constraining helium core WD models.

\end{abstract}


\section{INTRODUCTION}

	Extremely low mass (ELM) WDs, with masses $<$0.3 M$_{\odot}$, are the 
remnants of stars that never ignited helium in their cores. The Universe is not old 
enough to produce ELM WDs by single star evolution.  Thus ELM WDs must undergo 
significant mass loss during their evolution.  Although metal-rich red giants with 
strong winds may evolve in isolation into single $\simeq$0.4 \msun\ WDs 
\citep{kilic07c}, producing $\simeq$0.2 \msun\ ELM WDs most likely requires compact 
binary systems \citep[e.g.][]{marsh95}.  Observational data for ELM WDs is limited, 
however, because of their rarity.  For example, among the 9316 WDs identified in the 
Sloan Digital Sky Survey (SDSS), fewer than 0.2\% have masses below 0.3 M$_{\odot}$ 
\citep{eisenstein06, kepler07}.

	\citet{kilic09,kilic10} have established a radial velocity program to search 
for companions around known ELM WDs. Of the six ELM WDs observed to date -- 
including J0917+4638, the lowest mass WD known \citep{kilic07b} -- all six ELM WDs 
are in binaries with $\leq$1 day orbital periods. Two more recently identified ELM 
WDs, NLTT 11748 and J1257+5428 \citep{kawka09,marsh10,kulkarni10}, are also in 
binaries with $<$1 day orbital periods \citep{steinfadt10,kawka10,kilic10b}. Three 
of these eight ELM WDs will merge due to gravitational wave radiation in less than 
500 Myr \citep{mullally09, kilic10}.  A larger sample of ELM WDs is required to 
measure the space density, period distribution, and merger rate of these systems.

	Here we present 12 ELM WDs with $\le0.25$ \msun\ found in the Hypervelocity
Star (HVS) Survey of \citet{brown05, brown06, brown06b, brown07a, brown07b,
brown09a, brown09b}, ten of which are new discoveries.  Our ELM WD sample is unique
because it comes from a complete, non-kinematically-selected survey of stars
targeted in a well-defined range of apparent magnitude and color.  Radial velocity
follow-up reveals that 11 of the 12 ELM WDs are binaries with $<$14 hr orbital
periods.  Clearly, the compact binary formation scenario is the best explanation for
ELM WDs.

	In Section 2 we describe the survey design and the spectroscopic
observations.  In Section 3 we derive the physical parameters of the ELM WDs with
stellar atmosphere modeling, and present the radial velocity curves for each object.
In Section 4 we discuss the nature of the binary systems, and conclude in Section
5.

\begin{figure}		
 \plotone{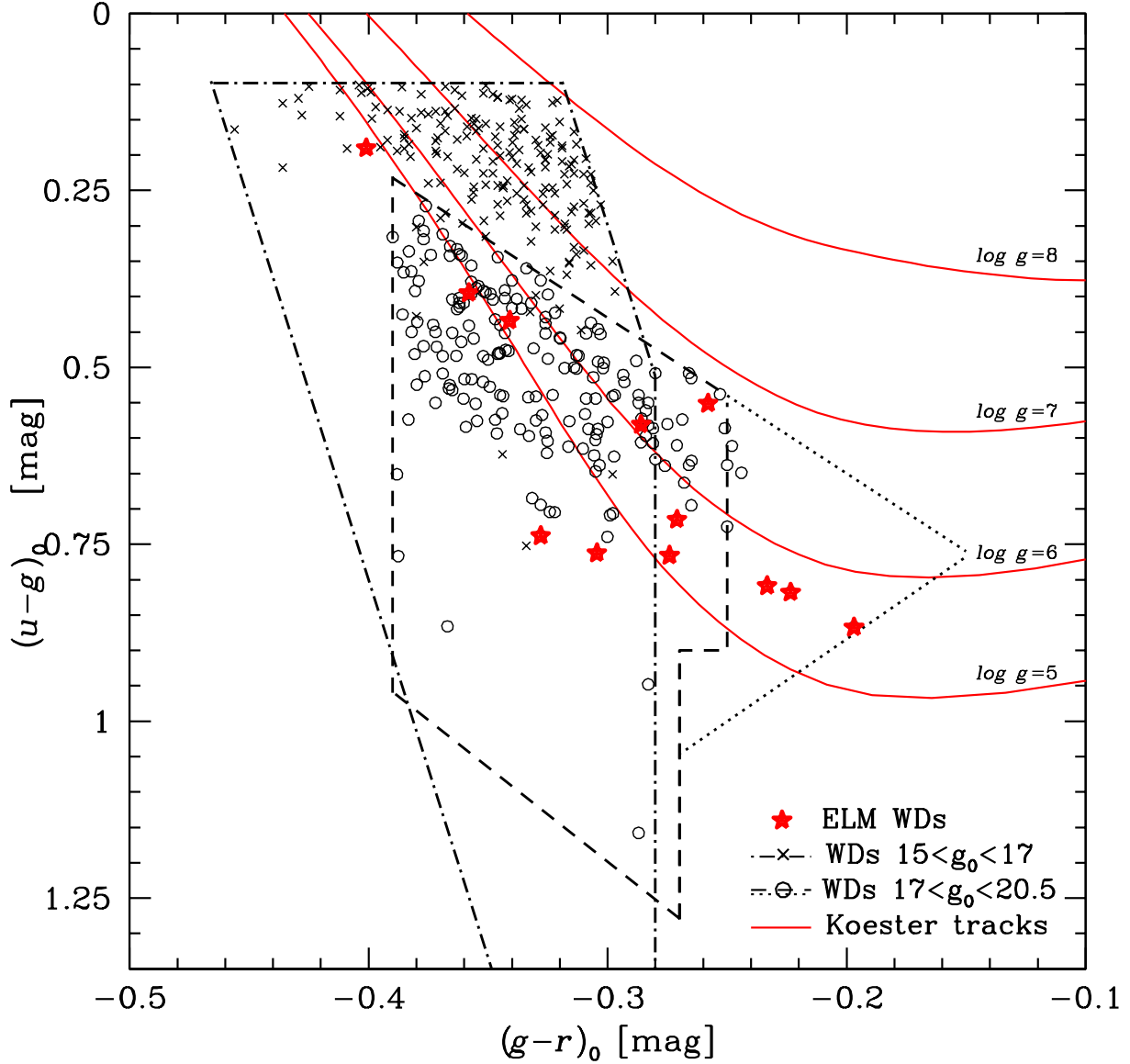}
 \caption{ \label{fig:ugr}
	Color-color diagram showing the observed distribution of ELM WDs ({\it solid
stars}) and the other WDs ({\it crosses, circles}) found in the HVS survey
(15$<$$g_0$$<$17 {\it dot-dash line}, 17$<$$g_0$$<$19.5 {\it dashed line},
19.5$<$$g_0$$<$20.5 {\it dotted line}) compared with our synthetic photometry of
the hydrogen atmosphere WD models of D. Koester ({\it solid lines}).
}
 \end{figure}

\section{DATA}

\subsection{Survey Design}

	The ELM WDs reported here are found in the HVS survey, a radial velocity
survey of objects with the colors of late-B type stars.  All survey targets were
selected from the Sloan Digital Sky Survey (SDSS) photometric catalog using
de-reddened, uber-calibrated PSF magnitudes \citep[][]{adelman08}.
	The HVS survey color selection is illustrated in Figure \ref{fig:ugr}.
Although the color selection was designed to exclude normal WDs by their $(u-g)$
color, the selection fortuitously includes low surface gravity WDs (see Figure
\ref{fig:ugr}).

	The HVS survey consists of three parts.  The first part of the HVS survey is 
defined by $15<g_0<17$ (dash-dot line in Fig.\ \ref{fig:ugr}) and is 100\% complete 
over 7300 deg$^2$ of the SDSS Data Release 5 footprint \citet{brown07a}.  The second 
part of the HVS survey is defined by $17<g_0<19.5$ (dashed line in Fig.\ 
\ref{fig:ugr}) and is 100\% complete over the same 7300 deg$^2$ of the SDSS Data 
Release 5 footprint \citet{brown07b}.  The final part extends the earlier surveys 
over 9800 deg$^2$ of the SDSS Data Release 7 footprint and also extends the 
magnitude limit to $g_0=20.5$ (dotted line), as described in \citet{brown09a}.  The 
final part of the HVS survey is currently 88\% complete.

	Our spectroscopy reveals that 15\% of the HVS Survey targets are WDs.  Two
of the WDs with $\simeq$0.2 \msun\ are previously published elsewhere
\citep{kilic07, kilic10}.  Here, we present the remaining ten WDs with mass
$\le$0.25 \msun .

\subsection{Spectroscopic Observations}

	With the exception of J0923+3028, we obtained all observations at the 6.5m
MMT telescope using the Blue Channel spectrograph.  We operate the spectrograph with
the 832 line mm$^{-1}$ grating in second order, providing wavelength coverage 3650
\AA\ to 4500 \AA\ and a spectral resolution of 1.0 - 1.2 \AA , depending on the slit
size used.  We obtain all observations at the parallactic angle, with a comparison
lamp exposure paired with every observation.

	We obtained spectroscopy for J0923+3028 at the Fred Lawrence Whipple
Observatory 1.5m Tillinghast telescope using the FAST spectrograph
\citep{fabricant98}.  We operate the FAST spectrograph with the 600 line mm$^{-1}$
grating, providing wavelength coverage 3600 \AA\ to 5500 \AA\ and a spectral
resolution of 2.3 \AA .  All observations are paired with a comparison lamp
exposure.

	We process the spectra using IRAF\footnote[2]{IRAF is distributed by the
National Optical Astronomy Observatories, which are operated by the Association of
Universities for Research in Astronomy, Inc., under cooperative agreement with the
National Science Foundation.} in the standard way.  We flux-calibrate using blue
spectrophotometric standards \citep{massey88}, and we measure radial velocities
using the cross-correlation package RVSAO \citep{kurtz98}.

\subsection{Radial Velocities}

	It is important to maximize velocity precision in our tests for variability, 
and we achieve the best precision by cross-correlating the ELM WDs with themselves. 
Our procedure begins by cross-correlating the observed spectra with a high 
signal-to-noise WD template to obtain preliminary velocities.  We then shift the 
spectra to the rest frame, and sum them together to create templates for each 
target.  Finally, we cross-correlate the spectra with the appropriate template to 
obtain the final velocities for each target.  The average radial velocity 
uncertainty of the measurements is $\pm18$ km s$^{-1}$.

	We check our velocities using WD model spectra with atmospheric parameters
customized for each target. The results are consistent within 10 km s$^{-1}$, which
we take as our systematic uncertainty.  Table \ref{tab:dat} in the Appendix presents
the full set of radial velocity measurements for the 10 newly discovered ELM WDs
presented here.

\begin{figure}		
 \plotone{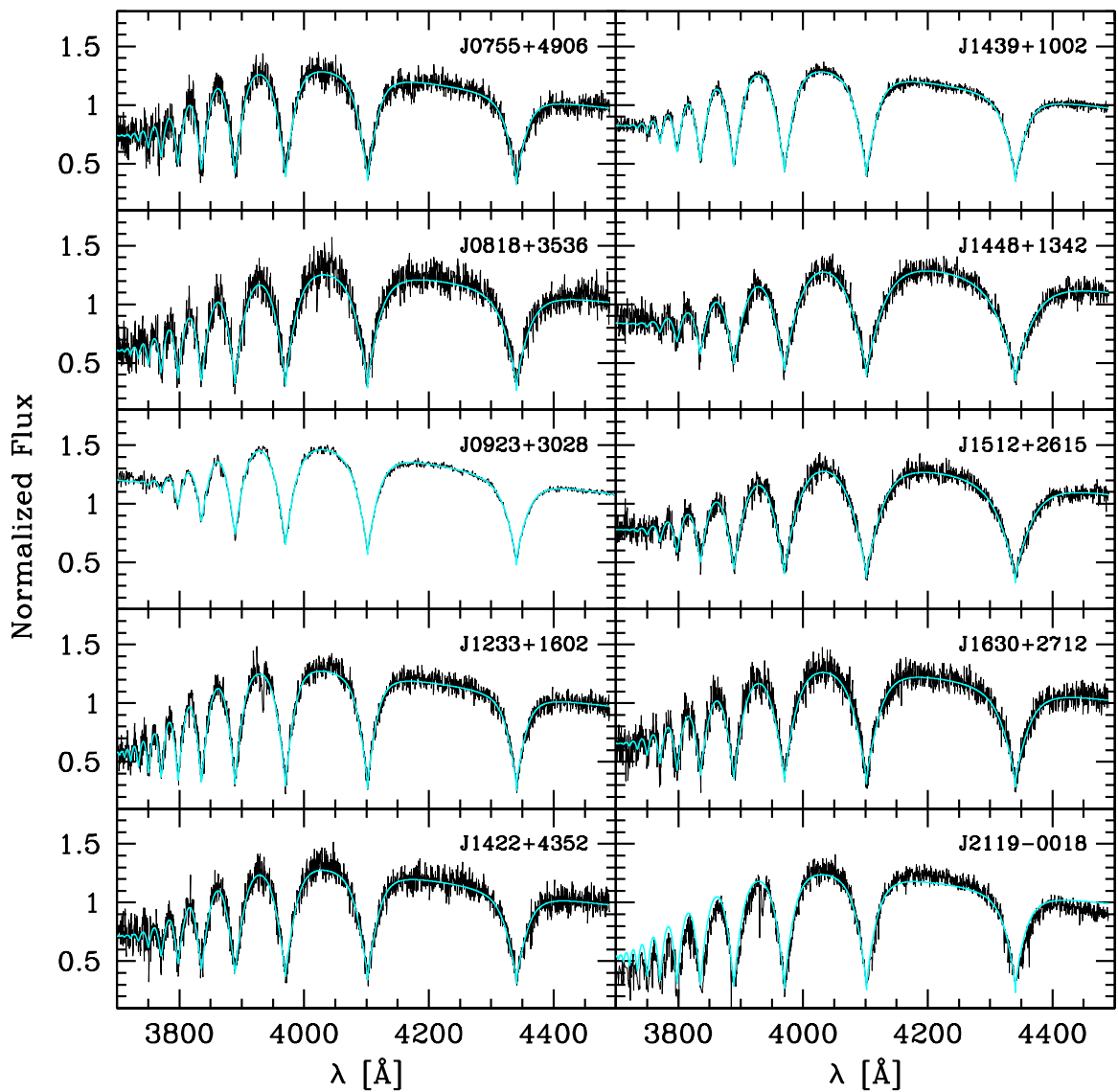}
 \caption{ \label{fig:spec}
	Model fits ({\it cyan lines}) overplotted on the composite observed spectra 
({\it black lines}).  The spectral continua provide improved $T_{\rm eff}$ 
determination except for the poorly fluxed-calibrated J2119$-$0018, for which we use 
the continuum-corrected spectrum. }
 \end{figure}

\begin{figure}		
 \plotone{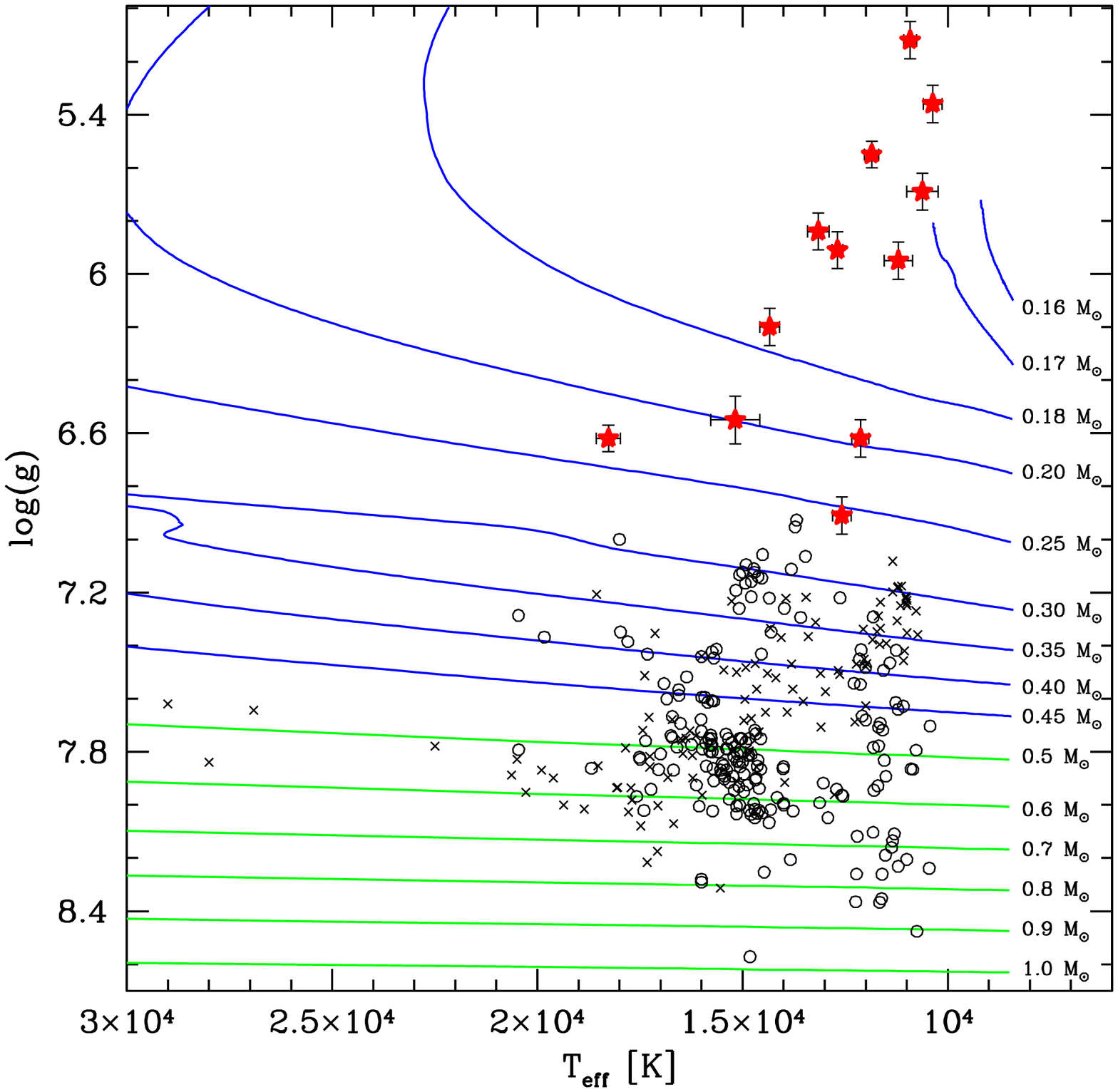}
 \caption{ \label{fig:teff}
	Surface gravity vs.\ effective temperature of the observed ELM WDs ({\it red 
stars}) and the other WDs ({\it crosses, circles}) found in the HVS survey, compared 
with predicted tracks for He WDs with 0.16--0.45 \msun\ \citep[{\it blue 
lines},][]{panei07} and CO WDs with 0.5--1.0 \msun\ \citep[{\it green 
lines},][]{bergeron95,holberg06}.  Our sample of ELM WDs is defined by $m\le0.25$ 
\msun . }
 \end{figure}

\section{RESULTS}

	Our time-series spectroscopy provides for robust determinations of effective 
temperature and surface gravity for each object as well as its binary orbital 
parameters.  Follow-up spectroscopy of J0917+4638 and J1053+5200 is already 
published \citep{kilic07b, kilic10}, but the other ELM WDs in our sample have not 
been studied until now.

\subsection{Stellar Atmosphere Parameters}

	We perform stellar atmosphere model fits using synthetic DA WD spectra
kindly provided by D.\ Koester.  The grid of WD model atmospheres covers effective
temperatures from 6000 K to 30,000 K in steps of 500 K to 2000 K, and surface
gravities from $\log{g}=$ 5.0 to 9.0 in steps of 0.25 dex.  The model atmospheres
are calculated assuming local thermodynamic equilibrium and include both convective
and radiative transport \citep{koester08}.
	We perform fits on the average composite spectra for each object and present
the resulting $T_{\rm eff}$ and $\log{g}$ values in Table \ref{tab:param}.  We also
perform fits to the individual spectra to derive a robust statistical error estimate
for each object, also presented in Table \ref{tab:param}.  Figure \ref{fig:spec}
shows the composite spectra and our model fits. The surface gravities range 5.1--6.9
dex at effective temperatures of 10400--18300 K, confirming that the objects are ELM
WDs.

	We fit the flux-calibrated spectral continua to better measure effective
temperature.  The exception is J2119$-$0018, the one ELM WD we observed under
variable conditions and at high airmass, for which we fit the continuum-corrected
spectrum.  If we fit only the continuum-corrected Balmer line profiles for all the
ELM WDs, we obtain best-fit solutions that differ by $440 \pm 190$ K in $T_{\rm
eff}$ and $0.04 \pm 0.04$ dex in $\log g$ from our published values.  These
differences reflect our systematic error, and demonstrate that our fits to the
entire flux-calibrated spectra are reasonably accurate.  Our fits also agree well
with the SDSS photometry in all five filters, an additional demonstration that our
temperature and surface gravity measurements are reliable.

	Figure \ref{fig:teff} compares the observed ELM WDs and other WDs in the HVS 
survey with the improved \citet{panei07} tracks \citep[see][]{kilic10} for He-core 
WDs with masses 0.16--0.45 \msun\ and the 
\citet{bergeron95}\footnote[3]{http://www.astro.umontreal.ca/$\sim$bergeron/CoolingModels/} 
tracks for normal CO-core WDs with masses 0.5--1 \msun . The gap between the 0.17 
\msun\ and 0.18 \msun\ He-core WD tracks is linked to the threshold for 
thermonuclear flashes in the hydrogen shell burning phase \citep{panei07}.  
Remarkably, the majority of observed ELM WDs fall in the gap of parameter space 
between the 0.17 \msun\ and 0.18 \msun\ models.

	The inconsistency between the observations and the He-core WD models make 
accurate mass and luminosity estimates difficult.  Fortunately, mass and luminosity 
change slowly over the range of effective temperature and surface gravity sampled by 
our ELM WDs.  We conclude that the majority of the ELM WDs have a mass of 
$\simeq$0.17 \msun\ and an absolute magnitudes of $M_g\simeq8$; more precise 
estimates are possible for the 0.20 - 0.25 \msun\ ELM WDs and are summarized in 
Table \ref{tab:param}.

	Using these absolute magnitude estimates, the majority of our ELM WDs are 
found at heliocentric distances of $1<d<3$ kpc.  The notable exception is 
J0923+3028, a bright $g=15.7$ ELM WD approximately 280 pc distant. All of our ELM 
WDs are located at high Galactic latitudes $30\arcdeg<|b|<80\arcdeg$ in the SDSS 
imaging footprint.

	The systemic velocities suggest that two objects are halo WDs: J0818+3536 
and J1422+4352 have systemic heliocentric radial velocities of $-201 \pm 4$ \kms\ 
and $-195 \pm 8$ \kms , respectively.  \citet{kilic10} also identify J1053+5200 as 
a halo object based on its proper motion and distance estimate.  Unfortunately, 
reliable proper motions are unavailable for our fainter $g\simeq20$ WDs.  Despite 
having relatively large $>1$ kpc distances above the Galactic plane, the remaining 
ELM WDs in our sample have systemic radial velocities consistent with a disk origin.

\subsection{Orbital Parameters}

\begin{figure}		
 \plotone{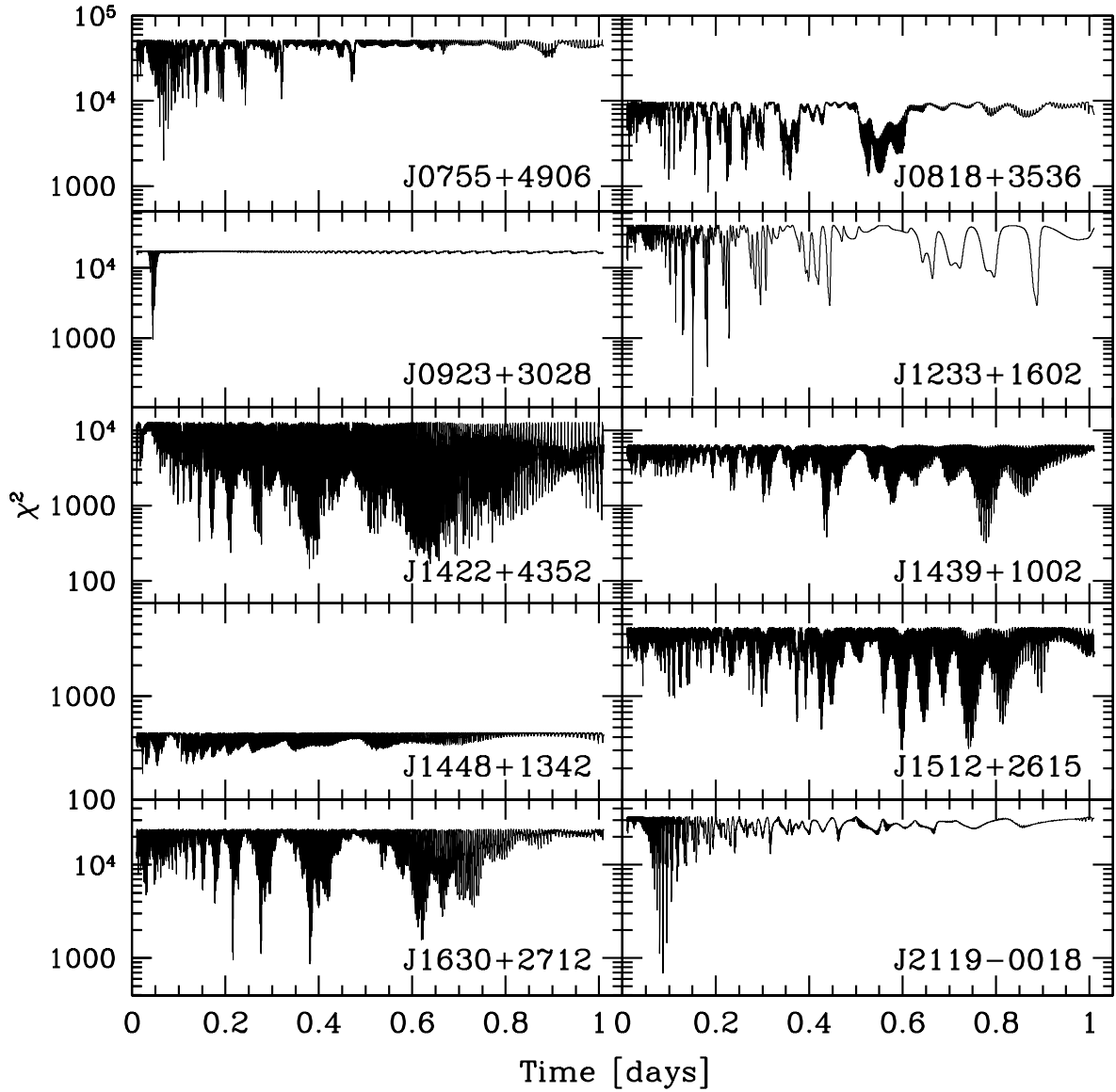}
 \caption{ \label{fig:pdm}
	Periodograms for the 10 new ELM WDs.  Some objects have multiple period
aliases because of insufficient coverage, however in all cases the periods are
constrained to $<$1 day.  J1448+1342 is formally consistent with no velocity
variability. }
 \end{figure}

	Eleven of our twelve ELM WDs exhibit significant radial velocity variation,
with peak-to-peak velocity amplitudes up to 890 \kms .  We compute orbital periods
for these systems by finding the period that minimizes $\chi^2$ for a circular
orbit.  Figure \ref{fig:pdm} plots the periodograms for the 10 new ELM WDs.  A few
ELM WDs have multiple period solutions because of insufficient coverage, however in
all cases the periods are constrained to be $<$1 day.  We estimate the period error
by conservatively identifying the range of periods with $\chi^2 \le 2 \chi^2_{\rm
min}$, where $\chi^2_{\rm min}$ is the minimum $\chi^2$.

	We compute best-fit orbital elements using the code of \citet{kenyon86},
which weights each velocity measurement by its associated error.  The uncertainties
in the orbital elements are derived from the covariance matrix and $\chi^2$.  To
verify these uncertainty estimates, we perform a Monte Carlo analysis where we
replace the measured radial velocity $v$ with $v + g \delta v$, where $\delta v$ is
the error in $v$ and $g$ is a Gaussian deviate with zero mean and unit variance.  
For each of 10000 sets of modified radial velocities, we repeat the periodogram
analysis and derive new orbital elements. We adopt the inter-quartile range in the
period and orbital elements as the uncertainty.  For binaries with multiple period
aliases, both approaches yield similar uncertainties.  When there are several
equally plausible periods, the Monte Carlo analysis selects all possible periods and
derives very large uncertainties.  In these cases, we adopt errors from the
covariance matrix for the lowest $\chi^2$ orbital period.

\begin{deluxetable}{lcccccccc}
\tabletypesize{\footnotesize}
\tablecolumns{9}
\tablewidth{0pt}
\tablecaption{White Dwarf Physical Parameters\label{tab:param}}
\tablehead{
\colhead{Object} &
\colhead{$g_0$} &
\colhead{$(u-g)_0$} &
\colhead{$(g-r)_0$}  &
\colhead{$T_{\rm eff}$} &
\colhead{$\log g$} &
\colhead{$M_g$} &
\colhead{$d_{helio}$} &
\colhead{Mass} \\
   & (mag) & (mag) & (mag) & (K) & & (mag) & (kpc) & (\msun )
}
	\startdata
J0755+4906    & $20.095 \pm 0.023$ & $0.763 \pm 0.086$ & $-0.304 \pm 0.039$ & $13160 \pm 260$ & $5.84 \pm 0.07$ & 8.0 & 2.62 & 0.17 \\
J0818+3536    & $20.484 \pm 0.030$ & $0.766 \pm 0.152$ & $-0.274 \pm 0.058$ & $10620 \pm 380$ & $5.69 \pm 0.07$ & 8.0 & 3.14 & 0.17 \\
J0917+4638$^1$& $18.696 \pm 0.019$ & $0.738 \pm 0.037$ & $-0.328 \pm 0.024$ & $11850 \pm 170$ & $5.55 \pm 0.05$ & 8.0 & 1.38 & 0.17 \\
J0923+3028    & $15.628 \pm 0.018$ & $0.190 \pm 0.027$ & $-0.401 \pm 0.021$ & $18350 \pm 290$ & $6.63 \pm 0.05$ & 8.4 & 0.28 & 0.23 \\
J1053+5200$^2$& $18.874 \pm 0.023$ & $0.395 \pm 0.049$ & $-0.358 \pm 0.045$ & $15180 \pm 600$ & $6.55 \pm 0.09$ & 8.7 & 1.08 & 0.20 \\
J1233+1602    & $19.829 \pm 0.018$ & $0.809 \pm 0.068$ & $-0.233 \pm 0.028$ & $10920 \pm 160$ & $5.12 \pm 0.07$ & 8.0 & 2.32 & 0.17 \\
J1422+4352    & $19.794 \pm 0.020$ & $0.715 \pm 0.061$ & $-0.271 \pm 0.032$ & $12690 \pm 130$ & $5.91 \pm 0.07$ & 8.0 & 2.29 & 0.17 \\
J1439+1002    & $17.812 \pm 0.016$ & $0.434 \pm 0.042$ & $-0.341 \pm 0.026$ & $14340 \pm 240$ & $6.20 \pm 0.07$ & 8.0 & 0.92 & 0.18 \\
J1448+1342    & $19.217 \pm 0.023$ & $0.581 \pm 0.047$ & $-0.286 \pm 0.033$ & $12580 \pm 230$ & $6.91 \pm 0.07$ & 9.9 & 0.73 & 0.25 \\
J1512+2615    & $19.241 \pm 0.021$ & $0.551 \pm 0.051$ & $-0.258 \pm 0.030$ & $12130 \pm 210$ & $6.62 \pm 0.07$ & 9.3 & 0.97 & 0.20 \\
J1630+2712    & $20.040 \pm 0.019$ & $0.818 \pm 0.083$ & $-0.223 \pm 0.031$ & $11200 \pm 350$ & $5.95 \pm 0.07$ & 8.0 & 2.56 & 0.17 \\
J2119$-$0018  & $20.000 \pm 0.021$ & $0.867 \pm 0.092$ & $-0.197 \pm 0.033$ & $10360 \pm 230$ & $5.36 \pm 0.07$ & 8.0 & 2.51 & 0.17 \\
	\enddata
\tablerefs{ $^1$ \citet{kilic07b}; $^2$ \citet{kilic10} }
\end{deluxetable}

\begin{deluxetable}{lclrcccc}
\tabletypesize{\footnotesize}
\tablecolumns{8}
\tablewidth{0pt}
\tablecaption{Binary Orbital Parameters\label{tab:orbit}}
\tablehead{
\colhead{Object}&
\colhead{$P$}&
\colhead{$K$}&
\colhead{$V_{\rm systemic}$}&
\colhead{Spec.\ Conjunction}&
\colhead{Mass Function}&
\colhead{$M_2$}&
\colhead{$\tau_{\rm merge}$}\\
  & (days) & (km s$^{-1}$) & (km s$^{-1}$) & (days + 2450000) & & (\msun ) & (Gyr)
}
	\startdata
J0755+4906    & $0.06302 \pm 0.00213$  & $438 \pm  5$ & $ -51 \pm  4$ &  $5150.84949 \pm 0.00010$ & $0.550 \pm 0.025$  & $\ge$ 0.81 & $\le$ 0.22 \\
J0818+3536    & $0.18315 \pm 0.02110$  & $170 \pm  5$ & $-201 \pm  4$ &  $5151.90466 \pm 0.00030$ & $0.094 \pm 0.014$  & $\ge$ 0.26 & $\le$ 8.89 \\
J0917+4638$^1$& $0.31642 \pm 0.00002$  & $145 \pm  2$ & $  28 \pm  1$ &  $3708.85755 \pm 0.00134$ & $0.102 \pm 0.003$  & $\ge$ 0.27 & $\le$ 36.8 \\
J0923+3028    & $0.04495 \pm 0.00049$  & $296 \pm  3$ & $   2 \pm  2$ &  $3818.67019 \pm 0.00002$ & $0.121 \pm 0.004$  & $\ge$ 0.34 & $\le$ 0.13 \\
J1053+5200$^2$& $0.04256 \pm 0.00002$  & $264 \pm  2$ & $  12 \pm  2$ &  $3790.79731 \pm 0.00004$ & $0.081 \pm 0.002$  & $\ge$ 0.26 & $\le$ 0.16 \\
J1233+1602    & $0.15090 \pm 0.00009$  & $336 \pm  4$ & $ -35 \pm  3$ &  $5268.84901 \pm 0.00014$ & $0.597 \pm 0.020$  & $\ge$ 0.86 & $\le$ 2.14 \\
J1422+4352    & $0.37930 \pm 0.01123$  & $176 \pm  6$ & $-195 \pm  8$ &  $4596.92484 \pm 0.00068$ & $0.208 \pm 0.023$  & $\ge$ 0.41 & $\le$ 42.9 \\
J1439+1002    & $0.43741 \pm 0.00169$  & $174 \pm  2$ & $ -55 \pm  1$ &  $3882.68848 \pm 0.00079$ & $0.240 \pm 0.007$  & $\ge$ 0.46 & $\le$ 54.7 \\
J1448+1342$^3$& \nodata                & $ 35 \pm  7$ & $ -28 \pm  5$ &  \nodata                  & \nodata            & \nodata    & \nodata    \\
J1512+2615    & $0.59999 \pm 0.02348$  & $115 \pm  4$ & $ -41 \pm  4$ &  $3879.48768 \pm 0.00119$ & $0.097 \pm 0.012$  & $\ge$ 0.28 & $\le$ 171  \\
J1630+2712    & $0.27646 \pm 0.00002$  & $218 \pm  5$ & $-118 \pm  3$ &  $5009.72779 \pm 0.00016$ & $0.295 \pm 0.018$  & $\ge$ 0.52 & $\le$ 15.6 \\
J2119$-$0018  & $0.08677 \pm 0.00004$  & $383 \pm  4$ & $ -28 \pm  4$ &  $5008.88701 \pm 0.00003$ & $0.501 \pm 0.016$  & $\ge$ 0.75 & $\le$ 0.54 \\
	\enddata
\tablerefs{ $^1$ \citet{kilic07b}; $^2$ \citet{kilic10} }
\tablecomments{ $^3$ The measurements are formally consistent with no variation.}
\end{deluxetable}

\clearpage

	We present the best-fit orbital parameters in Table \ref{tab:orbit}. Columns
include orbital period ($P$), radial velocity semi-amplitude ($K$), systemic
velocity ($V_{\rm systemic}$), the time of spectroscopic conjunction (the time when
the object passes through 0 \kms\ as it approaches the observer), mass function (see
Eqn.\ 1 below), and minimum secondary mass (assuming $i=90\arcdeg$).  The systemic
velocities in Table \ref{tab:orbit} are not corrected for the WDs' gravitational
redshifts, which should be subtracted from the observed velocities to find the true
systemic velocities. This correction is approximately 3 km s$^{-1}$ for our targets.

	We plot the best-fit orbits compared to the observed radial velocities in 
Figure \ref{fig:vel}, excluding the two objects previously published.  Follow-up 
observations were typically obtained over a pair of 3-day time baselines separated 
by one week, plus an original observation that extends the baseline up to 1488 days 
(not shown in the upper panel of Fig.\ \ref{fig:vel}).  The ELM WD's short orbital 
periods combined with our long time baselines allow us to constrain the orbital 
periods accurately.

\section{DISCUSSION}

\subsection{J0755+4906}

	SDSS J075552.40+490627.9 exhibits the largest radial velocity variation in 
our sample of ELM WDs.  The WD has a best-fit amplitude of $876\pm10$ \kms , an 
orbital period of $1.512\pm0.051$ hr, and a binary mass function of
	\begin{equation}
\frac{M_2^3~{\rm sin}^3i}{(M_1+M_2)^2}=\frac{P K^3}{2 \pi G}= 0.550 \pm 0.025 M_{\sun},
	\end{equation} where $i$ is the orbital inclination angle, $M_1$ is the 
ELM WD mass, and $M_2$ is the companion mass.  Although the inclination is unknown, 
we can use the distribution of possible inclinations to constrain the companion's 
mass.

	Given the observed orbital parameters, there is a 61\% probability that 
J0755+4906's companion is a WD with $<$1.4 \msun\ and a 20\% probability that the 
companion is a neutron star with 1.4-3.0 \msun .  The remaining probability is for a 
companion mass $>$3 \msun .  We estimate the most probable companion mass by 
assuming the mean inclination angle for a random stellar sample, $i=60\arcdeg$.  
For J0755+4906, the most likely companion is a 1.12 \msun\ object at an orbital 
separation of 0.7 \rsun .  Follow-up radio or X-ray observations are required to 
rule out a neutron star companion, but statistically the companion is most likely a 
massive WD.

	Short period binaries like J0755+4906 must eventually merge due to angular
momentum loss to gravitational wave radiation.  The merger time is
	\begin{equation}
\tau = \frac{(M_1 + M_2)^{1/3}}{M_1 M_2} P^{8/3} \times 10^{-2} {\rm Gyr}
	\end{equation} where the masses are in \msun\ and the period $P$ is in hours 
\citep{landau58}.  For a companion mass of 1.1 \msun , J0755+4906 will merge in 170 
Myr.  \citet{kilic10} discuss the many possible stellar evolution paths for such a 
merging system.  For J0755+4906, there is at least a 6\% likelihood that the system 
contains a pair of WDs whose total mass exceeds the Chandrasekhar mass and thus will 
explode as a Type Ia supernova. On the other hand, the extreme mass ratio 
$M_1/M_2\leq$0.21 means that mass transfer is likely to be stable and thus this 
binary will probably evolve into an AM CVn system.

\subsection{J0818+3536}

	The best-fit orbital period for SDSS J081822.34+353618.9 is $4.396 \pm 0.51$
hr, however the limited number of follow-up observations allow for aliases at 5.4
and 8.6 hr periods.  Periods longer than 14 hr are ruled out.  Thus J0818+3536 is a
short period binary.  Assuming the best-fit period, there is a 90\% probability that
the companion is a WD with $<$1.4 \msun\ and a 5\% probability that the companion is
a neutron star with 1.4-3.0 \msun . For the most probable inclination angle
$i=60\arcdeg$, the companion is a 0.33 \msun\ low mass WD at orbital separation of
1.1 \rsun .

\subsection{J0917+4638}

	Previously published by \citet{kilic07b}, SDSS J091709.55+463821.8 has a
binary orbital period of $7.5936\pm0.0005$ hr.  Assuming an inclination angle of
$60\arcdeg$, its companion is another low-mass WD with 0.35 \msun\ at an orbital
separation of 1.6 \rsun .

	Curiously, J0917+4638 and the two other ELM WDs with lower surface gravity 
in our sample (J1233+1602 and J2119$-$0018) all show significant Ca {\sc ii} K 
absorption in their atmospheres (Figure \ref{fig:spec}). The extremely short 
timescale for gravitational settling of Ca in the WD photospheres means there must 
be an external source for the observed metals \citep{koester06, kilic07}.  The ELM 
WDs are located far above the Galactic plane where accretion from the interstellar 
medium is unlikely.  A plausible source of Ca is accretion from a circumbinary disk 
created during the mass loss phase of the giant.  Since the ELM WDs all went through 
a common envelope phase with a companion, a left-over circumbinary disk is possible.

\subsection{J0923+3028}

	SDSS J092345.60+302805.0 is the brightest ELM WD in our sample, with $g=15.7$ 
mag and a heliocentric distance estimate of 280 pc.  J0923+3028 has a binary orbital 
period of $1.079 \pm 0.012$ hr and a mass function of $0.121 \pm 0.004$ \msun . 
There is a 87\% probability that the companion is a WD with $<$1.4 \msun\ and a 6\% 
probability that the companion is a neutron star with 1.4-3.0 \msun . For the most 
probable inclination angle, $i=60\arcdeg$, the companion is a 0.44 \msun\ WD at 
orbital separation of 0.5 \rsun . Thus, J0923+3028 is the second shortest period 
double WD system after J1053+5200 \citep{kilic10}. This system will merge in less 
than 130 Myr.

	J0923+3028 is the one ELM WD in our sample with a significant proper motion 
measurement, $\mu_{\alpha} cos \delta = -4.2 \pm 3.5$ and $\mu_{\delta}=-25 \pm 3.5$ 
mas yr$^{-1}$ \citep{munn04}.  The gravitational redshift of the WD is $\simeq$4 
\kms , thus its true systemic velocity is $-2\pm5$ \kms .  The velocity components 
with respect to the local standard of rest \citep{schonrich10} are $U=14 \pm 6$, 
$V=-20 \pm8$, $W=-2 \pm 6$ \kms .  J0923+3028 is clearly a disk star.

\subsection{J1053+5200}

	Previously published by \citet{kilic10}, SDSS J105353.89+520031.0 has a
binary orbital period of $1.0224\pm0.0005$ hr.  Assuming an inclination angle of
$60\arcdeg$, the companion is another low mass WD with 0.33 \msun\ at an orbital
separation of 0.4 \rsun .  This system will merge in less than 160 Myr.

\subsection{J1233+1602}

	SDSS J123316.20+160204.6 is the lowest surface gravity WD known.  Its
surface gravity, $\log{g}=5.12\pm0.07$, is comparable to a 2.5 \msun\ main sequence
star with $\log{g}\simeq4.2$ \citep{girardi02, girardi04}.  However, J1233+1602's
binary orbital parameters exclude the possibility of it being a main sequence star.

	J1233+1602 is a binary with a $3.6216\pm0.0022$ hr orbital period and a mass
function of $0.597 \pm 0.020$ \msun .  By assuming an edge-on orbit, $i=90\arcdeg$,
we can place a lower limit on the companion mass. For a 2.5 \msun\ A star primary,
the star must have at least a 2.44 \msun\ companion at an orbital separation of 2.0
\rsun .  However, the radius of a 2.5 \msun\ main sequence star is 1.9 \rsun . Thus
the orbital separation of two A stars in this system is less than their summed
radii.  Even if the companion were a 2.44 \msun\ neutron star, the system would
still be in Roche-lobe overflow, for which we see no evidence.

	As a check, we combine the spectra near maximum blue-shift and red-shift
into two composite spectra. If there is a contribution from the companion, it may be
visible as second pair of lines or an asymmetry in the line profiles.  We do not see
any extra lines or asymmetries in the line profiles.  We conclude J1233+1602 is a
$\simeq$0.17 \msun\ WD orbiting a much fainter companion.

	Given the observed mass function, there is a 58\% probability that the
companion is a WD with $<$1.4 \msun\ and a 21\% probability that the companion is a
neutron star with 1.4-3.0 \msun . Assuming the most probable orbital inclination
angle of $60\arcdeg$, J1233+1602's companion is a 1.2 \msun\ WD at an orbital
separation of 1.3 \rsun .  Follow-up radio or X-ray observations are required to
rule out a neutron star companion, but statistically the companion is most likely a
massive WD.

	There is a 7\% likelihood that this system contains a pair of WDs whose
total mass exceeds the Chandrasekhar mass and thus will explode as a Type Ia
supernova.  The merger time for such a system is $<$2 Gyr.  However, the extreme
mass ratio $M_1/M_2\leq0.20$ suggests that mass transfer will be stable and this
binary will likely evolve into an AM CVn system.

\subsection{J1422+4352}

	SDSS J142200.74+435253.2 has the fewest follow-up observations yet its
binary orbital period is reasonably well constrained.  The best-fit period is
$9.10\pm0.27$ hrs, with a significant alias at 15.3 hrs.  Assuming the best-fit
period, there is a 82\% probability that the companion is a WD with $<$1.4 \msun\
and a 8\% probability that the companion is a neutron star with 1.4-3.0 \msun . For
the most probable inclination angle, $i=60\arcdeg$, the companion is a normal 0.55
\msun\ WD at orbital separation of 2.0 \rsun .

\subsection{J1439+1002}

	SDSS J143948.40+100221.7 is a binary with a best-fit period of $10.498 \pm
0.041$ hrs.  Despite having follow-up observations on 10 nights spaced over one year
(see Fig.\ \ref{fig:vel}) there remains a significant period alias at $18.576 \pm
0.215$ hr.  Using the best-fit orbital solution, there is a 80\% probability that
the companion is a WD with $<$1.4 \msun\ and a 10\% probability that the companion
is a neutron star with 1.4-3.0 \msun . For the most probable inclination angle,
$i=60\arcdeg$, the companion is a normal 0.62 \msun\ WD at orbital separation of 2.2
\rsun .

\subsection{J1448+1342}

	SDSS J144801.13+134232.8 is the one ELM WD for which we detect no
significant velocity variation.  The best-fit orbit (0.5 hr period) has a reduced
$\chi^2$ identical to a constant-velocity fit, thus this system can have a velocity
semi-amplitude no larger than $K\le35\pm11$ \kms .  By comparison, the average
semi-amplitude of the other eleven ELM WD binaries is 240 \kms .  The required
orbital inclination for J1448+1342 to have the average semi-amplitude of the other
ELM WD binaries is $i\le8\arcdeg$.

	J1448+1342 may be either a pole-on binary system or a single star.  If the
orbital inclinations of our non-kinematically selected ELM WDs are randomly
distributed, we expect one of the twelve ELM WDs to have $i\le23.5\arcdeg$; there is
a 12\% likelihood of finding one of the twelve systems with $i\le8\arcdeg$.  Thus it
is possible that J1448+1342 is a pole-on binary system.  The alternative is that
J1448+1342 is a single star.  It is intriguing that J1448+1342 is the most massive
object in our sample of 12 ELM WDs.  Existing stellar evolution models, however,
cannot explain 0.25 \msun\ WDs from single star evolution, even with extreme mass
loss models \citep{kilic07c}.  Additional observations are required to determine the
nature of this ELM WD. 

\subsection{J1512+2615}

	SDSS J151225.70+261538.5 is a binary with a best-fit orbital period of
$14.40 \pm 0.56$ hr, and a significant period alias at 17.8 hrs.  Assuming the
best-fit period, there is an 89\% probability that the companion is a WD with $<$1.4
\msun\ and a 5\% probability that the companion is a neutron star with 1.4-3.0 \msun
. For the most probable inclination angle $i=60\arcdeg$, the companion is a 0.36
\msun\ low mass WD at orbital separation of 2.5 \rsun .

\subsection{J1630+2712}

	SDSS J163026.09+271226.5 has a best-fit orbital period of $6.6350 \pm 
0.0005$ hr.  The observations also allow for 5.16 and 9.17 hr periods. Assuming the 
best-fit period, there is a 77\% probability that the companion is a WD with $<$1.4 
\msun\ and a 11\% probability that the companion is a neutron star with 1.4-3.0 
\msun . For the most probable inclination angle $i=60\arcdeg$, the companion is a normal 
0.70 \msun\ WD at an orbital separation of 1.7 \rsun .

\subsection{J2119$-$0018}

	SDSS J211921.96$-$001825.8 is the second lowest surface gravity WD known, 
with $\log{g}=5.36\pm0.07$.  As with J1233+1602, J2119$-$0018's orbital parameters 
exclude the possibility of it being a main sequence A star.

	J2119$-$0018 is a binary with a $2.0825\pm0.0010$ hr orbital period and a 
mass function of $0.501\pm0.016$ \msun .  If the primary is 2.5 \msun\ main sequence 
star, its companion must be $\ge$2.2 \msun\ at an orbital separation of 1.4 \rsun .  
This separation is smaller than the radius of main sequence stars.  We do not see 
any evidence of a companion in our spectra.  Thus J2119$-$0018 must be a 
$\simeq$0.17 \msun\ WD.

	Given the observed orbital parameters, there is a 64\% probability that the 
companion is a WD with $<$1.4 \msun\ and a 18\% probability that the companion is a 
neutron star with 1.4-3.0 \msun .  Assuming the most probable orbital inclination 
angle of $60\arcdeg$, J2119$-$0018's companion is a 1.04 \msun\ WD at an orbital 
separation of 0.9 \rsun .  Follow-up radio or X-ray observations are required to 
rule out a neutron star companion, but the companion is most likely a massive WD.

	This system will merge in less than 540 Myr.  There is a 5\% likelihood that
the system contains a pair of WDs whose total mass exceeds the Chandrasekhar mass
and thus will explode as a Type Ia supernova.  However, the extreme mass ratio
$M_1/M_2\leq0.23$ suggests that mass transfer will be stable and this binary will
likely evolve into an AM CVn system.

\section{CONCLUSION}

	We present a complete sample of 12 ELM WDs with masses $\le0.25$ \msun\ in
the HVS Survey.  Eleven of the WDs are binary systems with orbital periods $<$14 hr;  
J1448+1342 is the only WD for which we detect no velocity variation.  Given the
binary nature of the other WDs in our non-kinematically selected sample, J1448+1342
is quite possibly a pole-on system.  Our observations demonstrate that the formation
of $\le$0.25 \msun\ ELM WDs requires close binary evolution \citep{marsh95}.

	Two of our targets, J1233+1602 and J2119$-$0018, have surface gravities 
lower than the previous record-holder J0917+4638 \citep[][]{kilic07} and thus are 
the lowest surface gravity WDs currently known.
	Interestingly, all three of these low surface gravity ELM WDs exhibit strong
Ca absorption in their spectra.  The Ca is likely explained by on-going accretion
from circumbinary disks.  Other sources of accretion are much less likely, given the
extremely short timescale for gravitational settling in the ELM WD photospheres and
their 1-2 kpc distances above the disk plane.

	One possible source of circumbinary disk accretion is debris from former
planetary systems \citep{dong10}.  For the case of our ELM WD binaries, the systems
must have had circumbinary planets that could survive the binary common envelope
evolution.  Common envelope evolution may also lead to ``second generation'' planet
formation \citep{perets10b}, the debris of which may be accreting onto the ELM WDs.  
In any case, the lowest surface gravity ELM WDs are interesting targets for future
infrared observations.

	Six of our 12 ELM WDs (50\%) will merge within a Hubble time (we exclude
J1448+1342 due to its unknown period); four of these mergers will happen in $\le$600
Myr.  In all cases the most likely binary companions are WDs, although neutron stars
cannot be ruled out based on the available data.  Two of the merger systems,
J0818+3536 and J1053+5200, are likely to have unusually low-mass ($<$0.4 \msun ) WD
companions and thus are possible progenitors for single helium-enriched sdO stars
\citep{heber09}.  Four of the merger systems are likely to have C/O WD companions
and thus will likely form extreme helium stars.  However, three of the systems
(J0755+4906, J1233+1602, and J2119$-$0018) have such extreme mass ratios $\leq$0.23
that future mass transfer is likely to be stable and thus they will evolve into AM
CVn systems.

	We expect that at least one of our ELM WDs has $i>85\arcdeg$ and thus is an 
eclipsing system.  \citet{steinfadt10} recently discovered the first eclipsing 
detached double WD; finding eclipsing WD binaries in our sample will provide 
important mass-radius constraints on He core ELM WD models.  The most probable 
eclipsing candidates in our sample are the systems with the shortest merger times, 
namely J0923+3028, J1053+5200, J0755+4906, and J2119$-$0018.

	Our sample of 12 ELM WDs is unique in terms of its non-kinematic selection
and its completeness.  Previous discoveries of ELM WDs in the SDSS \citep{liebert04,
eisenstein06} and other surveys \citep{kawka06, kawka09} have been useful for
studying the formation and future evolution of the individual systems
\citep{kilic10}.  However, the spectroscopic selection biases present in the SDSS
prohibit an accurate estimate of the space density of these objects.  SDSS WDs were
observed by different targeting programs selected with different color, magnitude, and
photometric quality cuts as well as different sparse sampling rates that vary over
the SDSS survey region \citep{eisenstein06}.
	Our sample, on the other hand, is complete in magnitude, color, and spatial
coverage and enables us to estimate the space density and merger rate of ELM WDs in
the Milky Way.  We explore possible links to underluminous supernovae, AM CVn stars,
and other phenomena in a companion paper \citep{brown10d}.

\acknowledgments
	We thank M.\ Alegria, J.\ McAfee, and A.\ Milone for their assistance with
observations obtained at the MMT Observatory, and P.\ Berlind and M.\ Calkins for
their assistance with observations obtained at the Fred Lawrence Whipple
Observatory.  We especially thank P.\ Challis obtaining additional observations of
J2119$-$0018, and the referee for helpful comments.  This project makes use of data
products from the Sloan Digital Sky Survey, which is managed by the Astrophysical
Research Consortium for the Participating Institutions.  This research makes use of
NASA's Astrophysics Data System Bibliographic Services.  To perform the Monte Carlo
analysis, we used the routine randgen.f, written by R.\ Chandler and P.\ Northrop.  
This work was supported in part by the Smithsonian Institution.  MK is supported by
NASA through the {\em Spitzer Space Telescope} Fellowship Program, under an award
from CalTech.

{\it Facilities:} \facility{MMT (Blue Channel Spectrograph)}

\begin{figure}		
 \plottwo{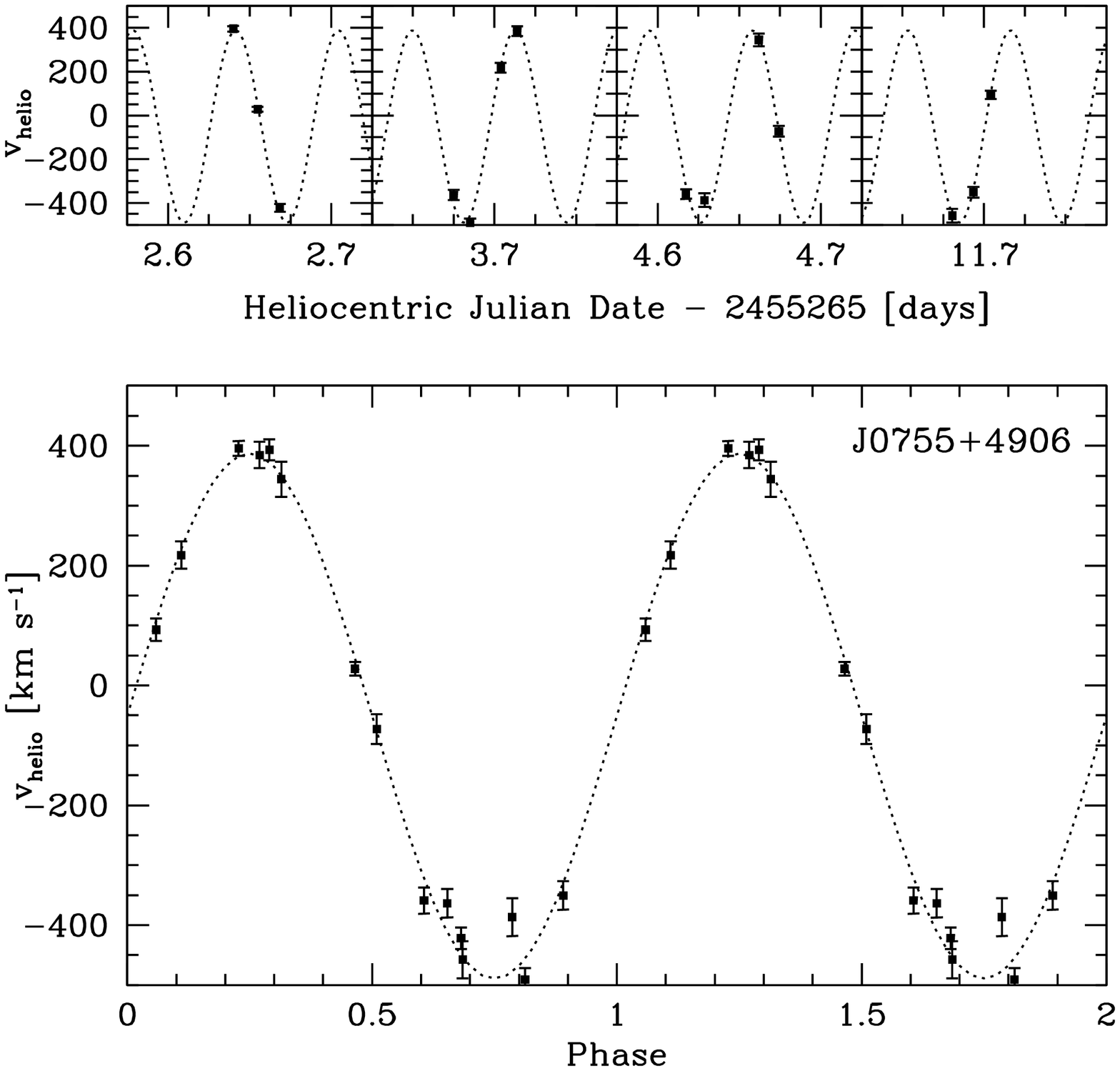}{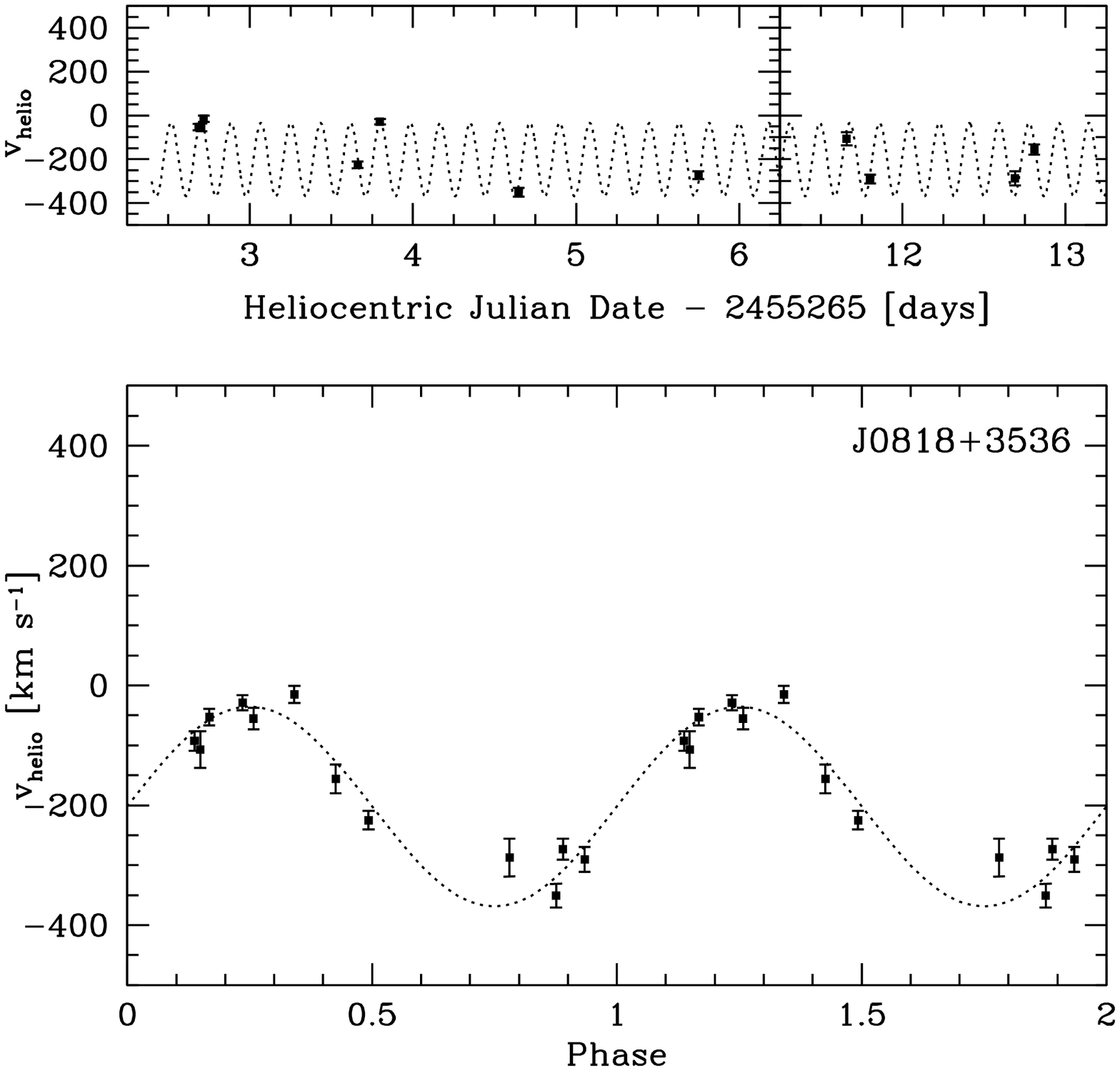}
 \plottwo{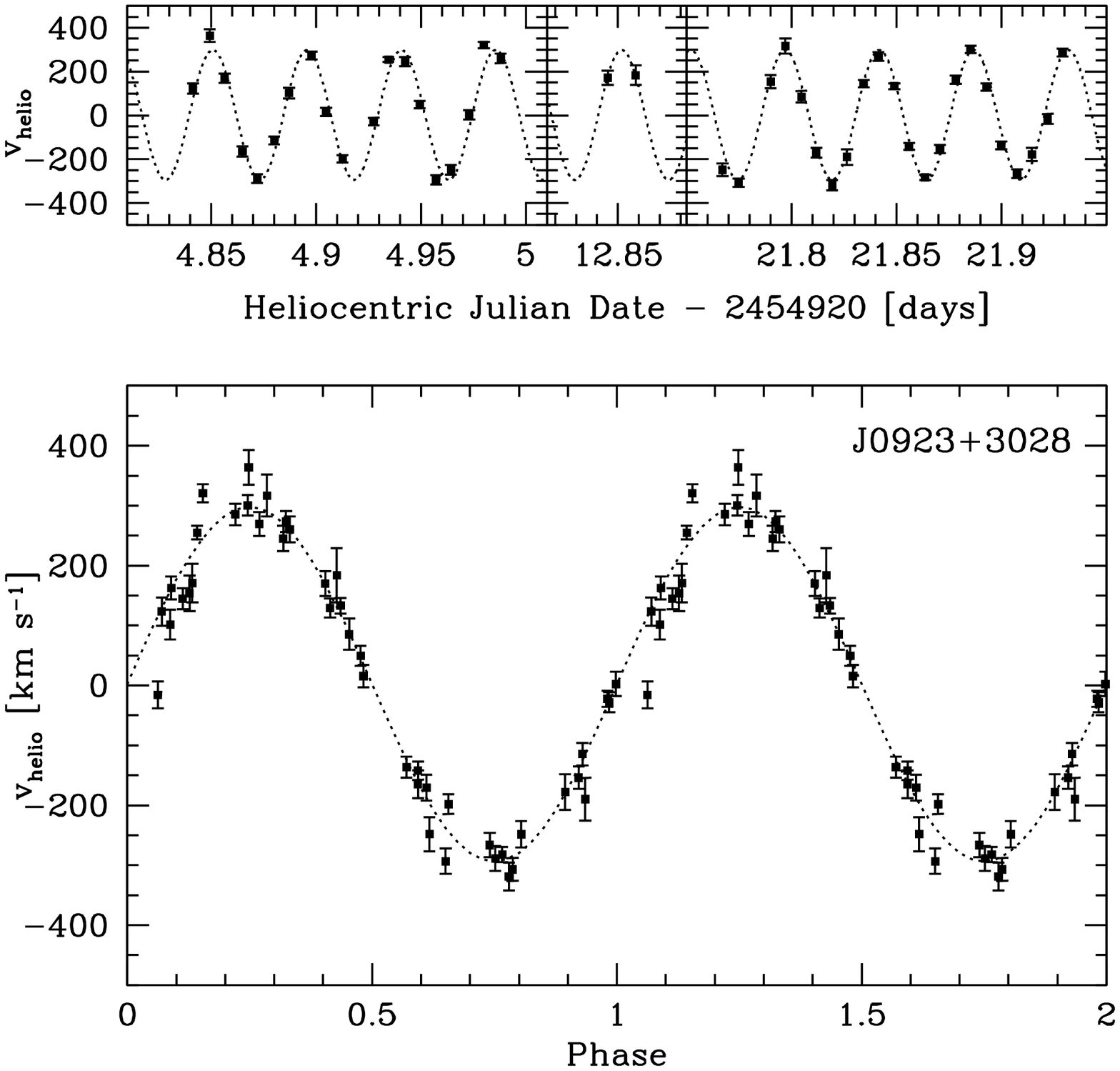}{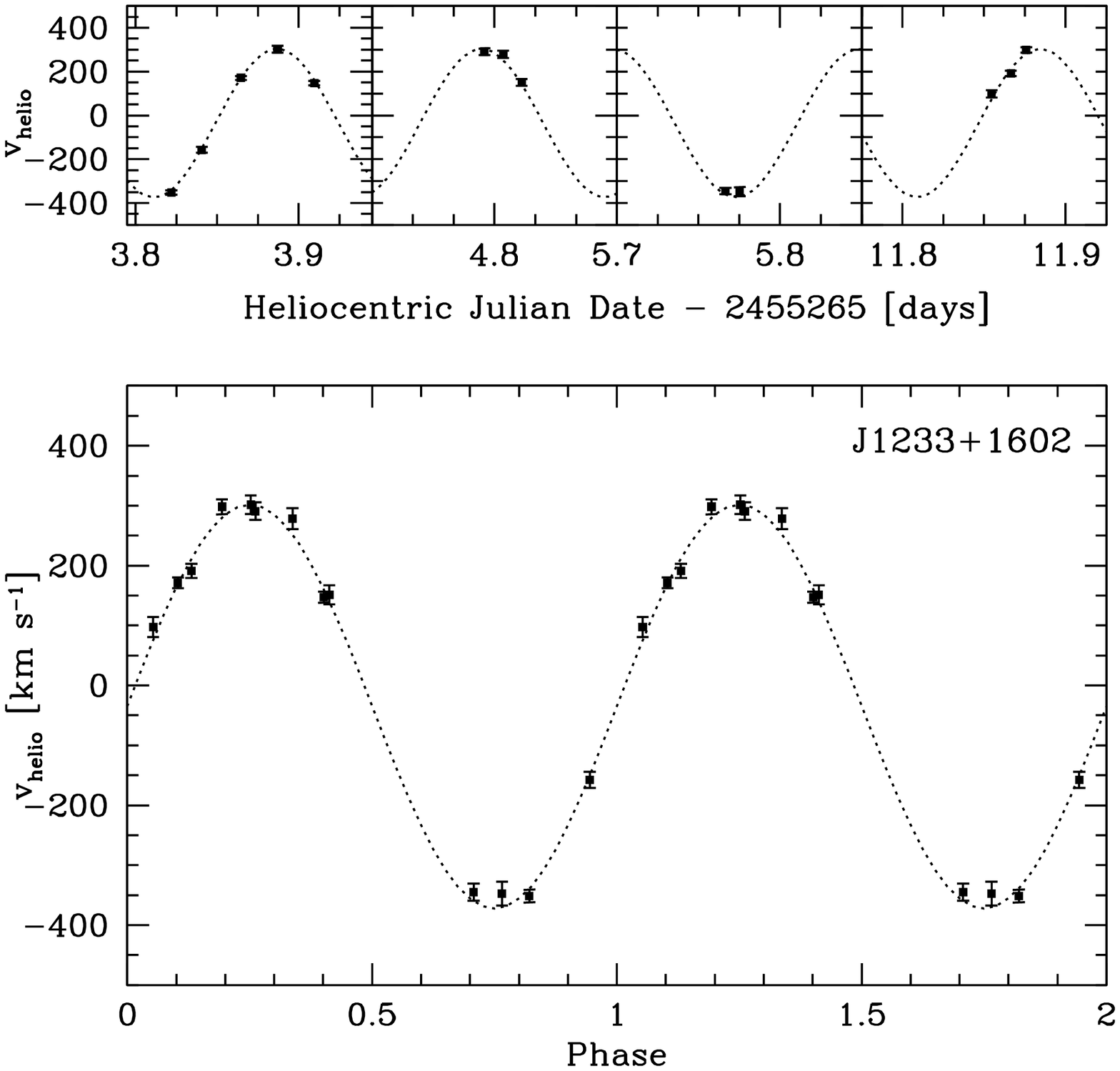}
 \caption{ \label{fig:vel}
	Data and best-fit orbits for the 10 new ELM WDs.  Upper panels plot
the heliocentric radial velocities vs.\ time.   Lower panels plot the observations
phased to the best-fit orbital solution (Table \ref{tab:orbit}), and also include
the observation first used to identify the ELM WD candidate.  The same vertical
axis is used in all 10 plots.
}
 \end{figure}

\begin{figure} \figurenum{5}
 \plottwo{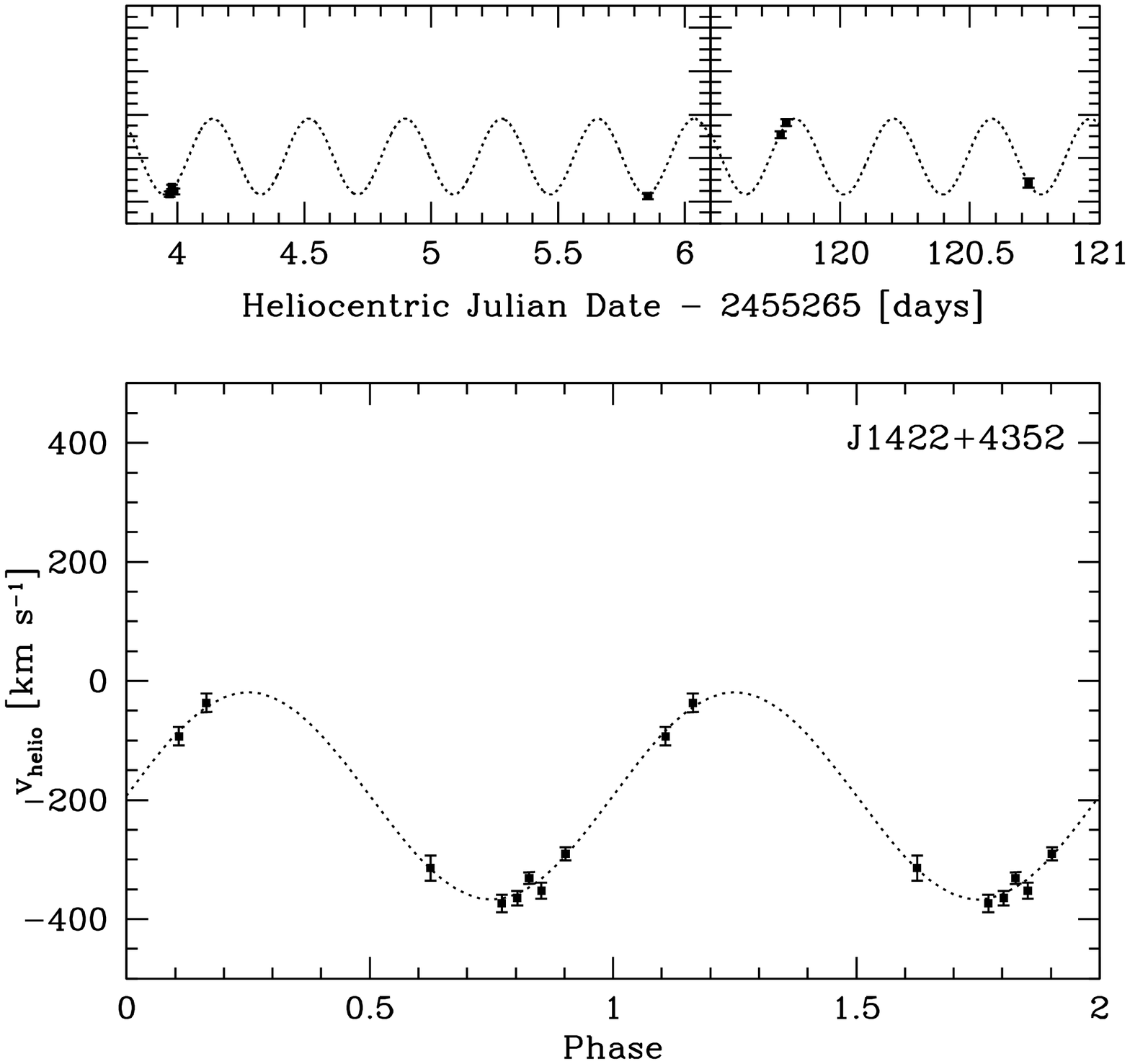}{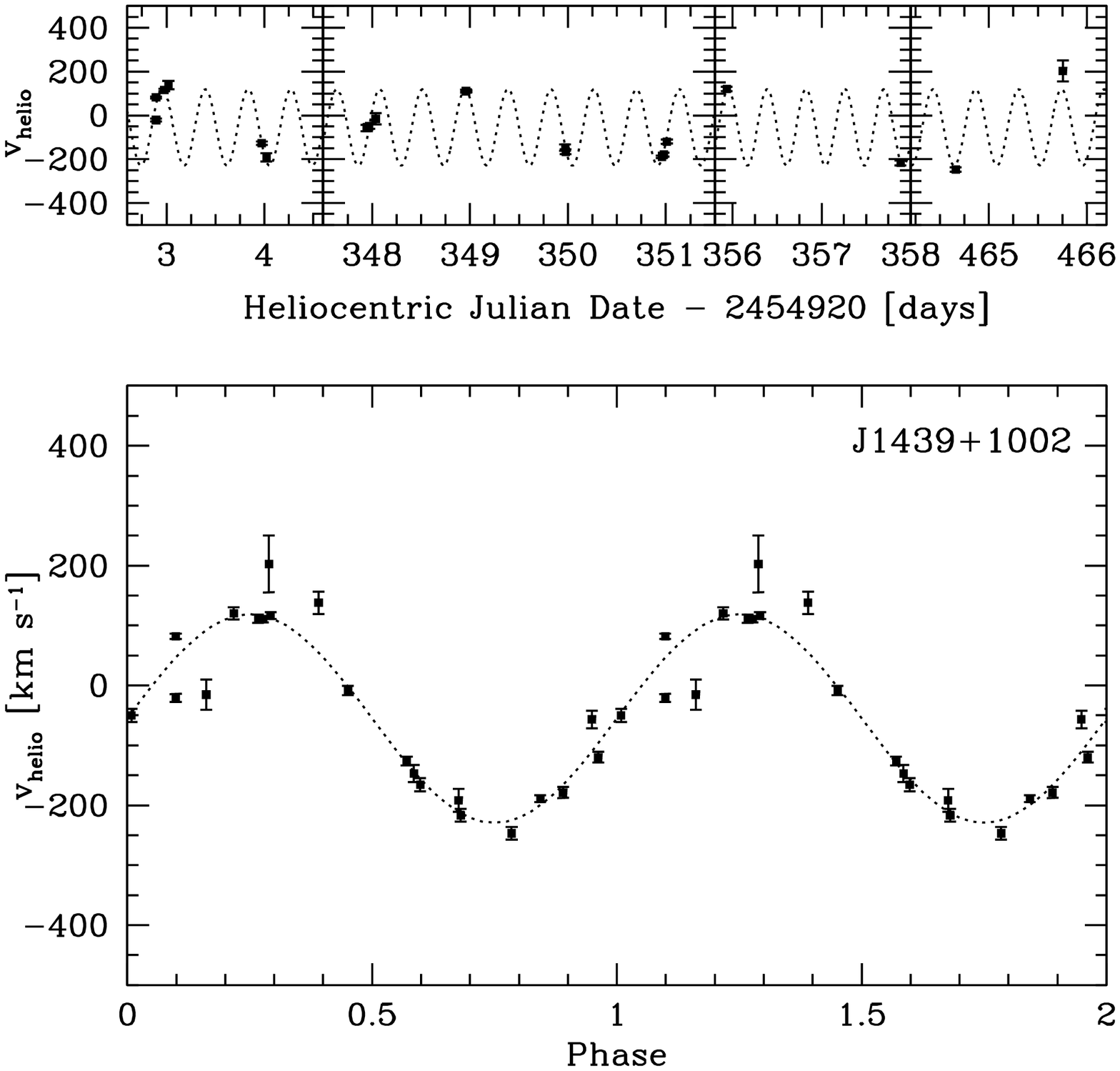}
 \plottwo{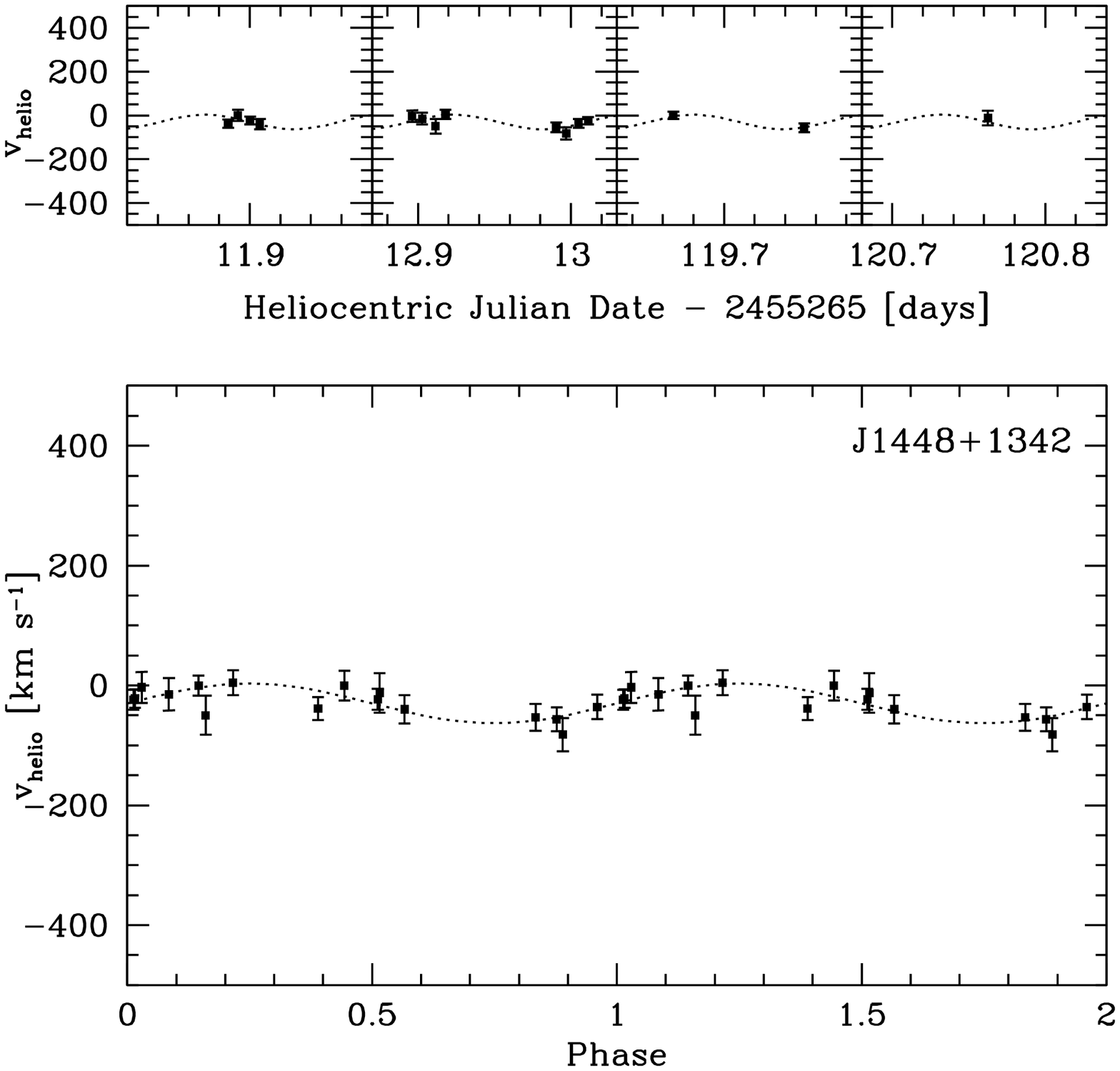}{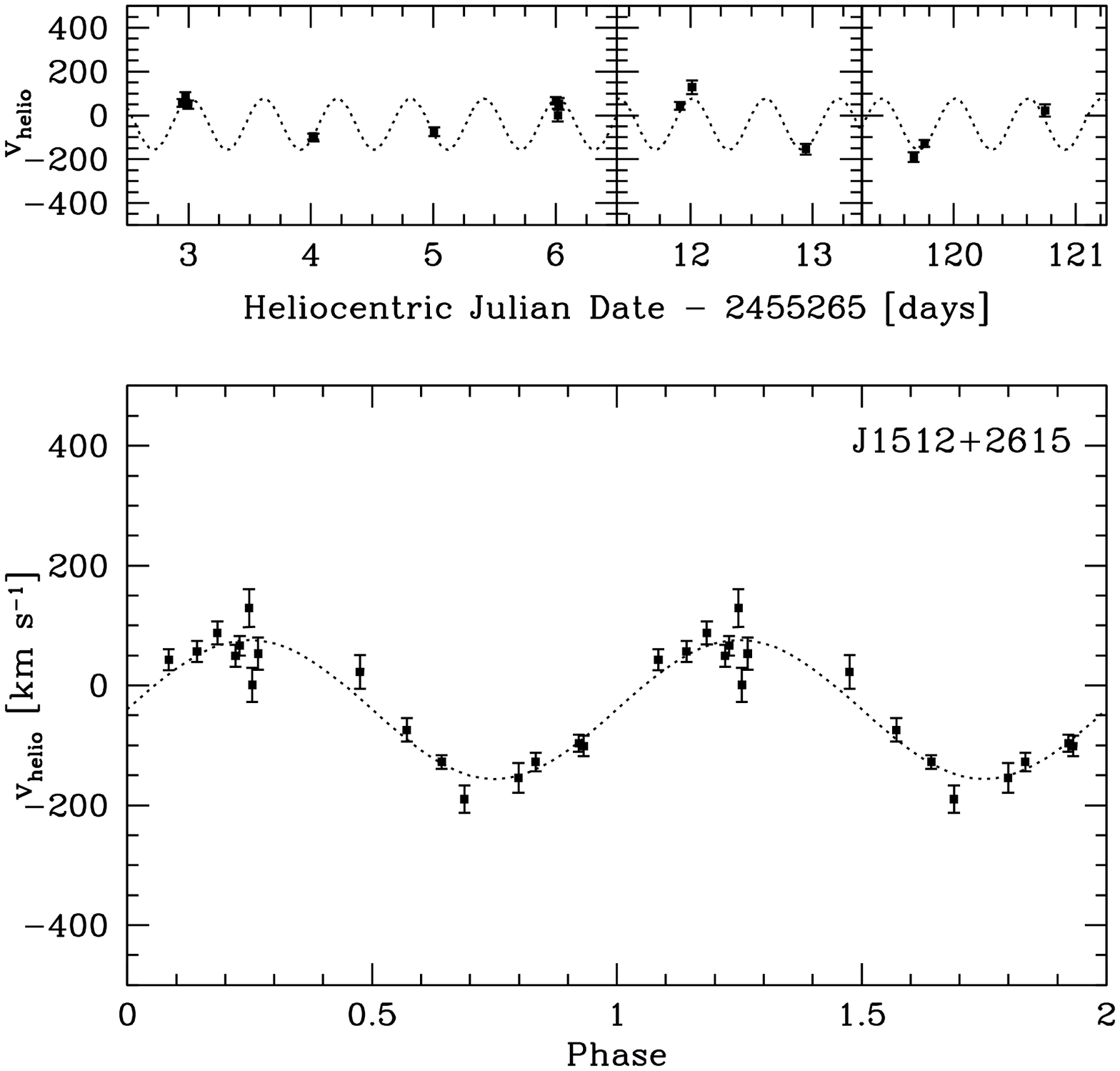}
\caption{Continued}
\end{figure}

\begin{figure} \figurenum{5}
 \plottwo{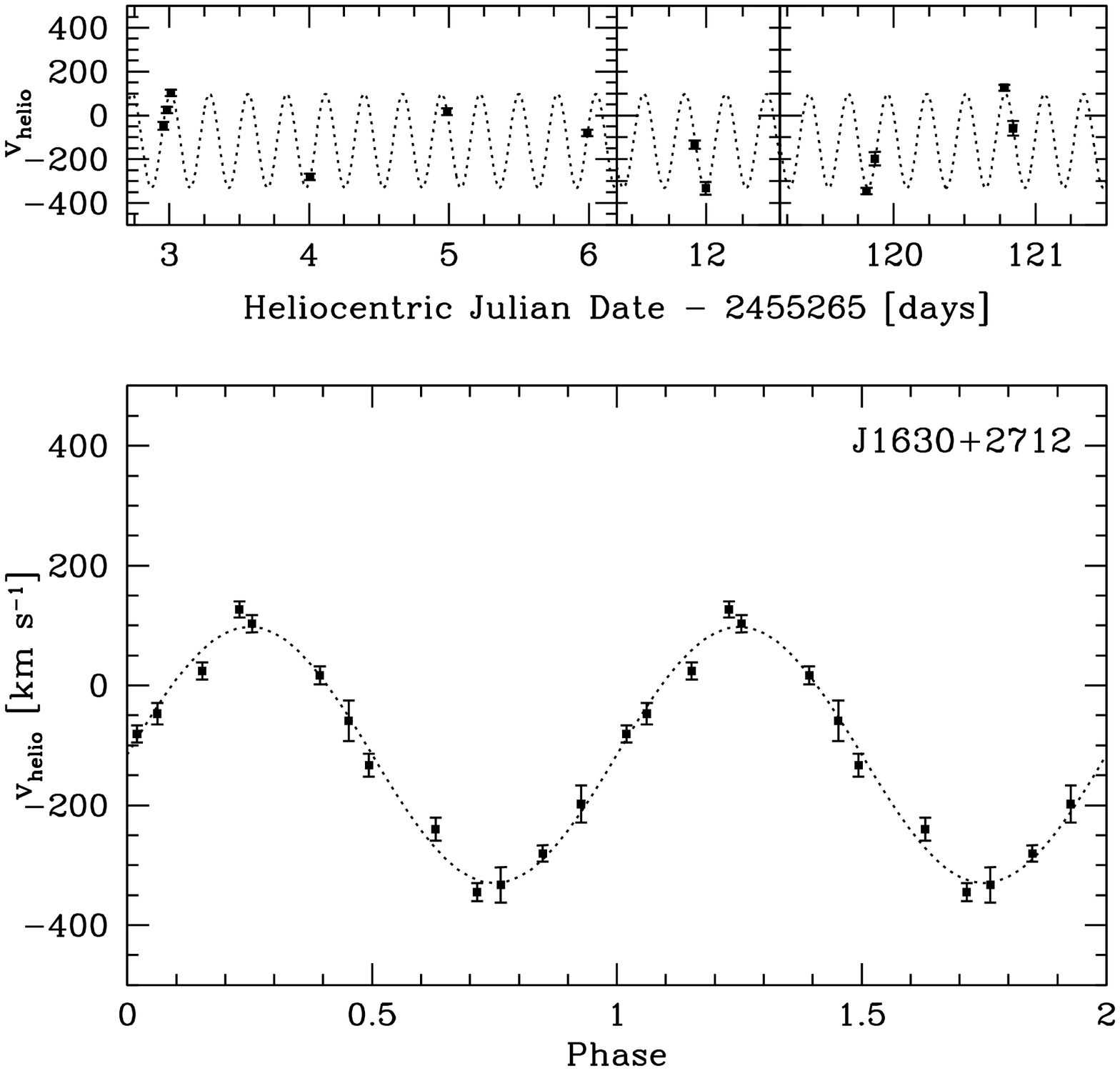}{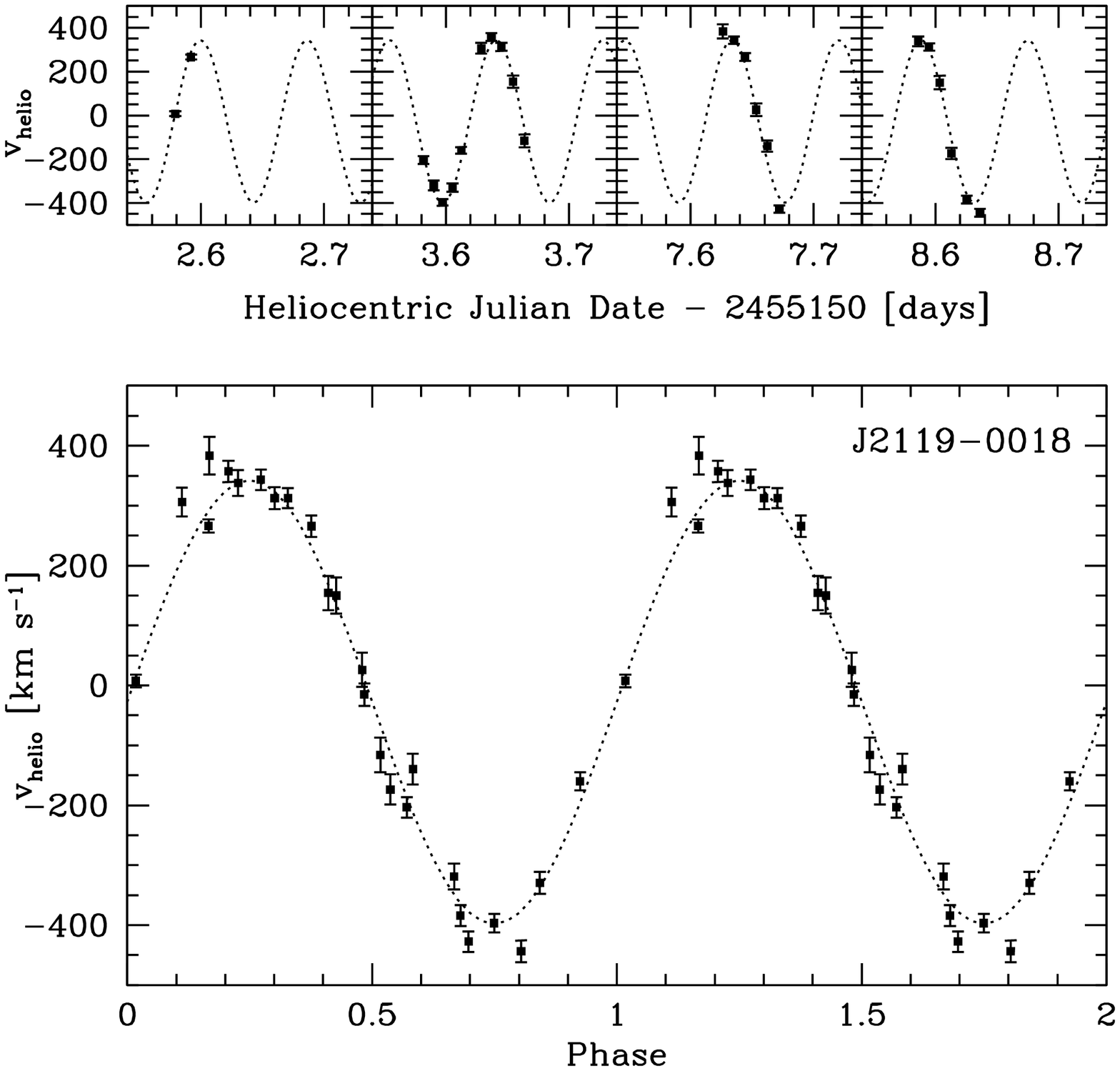}
\caption{Continued}
\end{figure}

\clearpage

\begin{thebibliography}{43}
\expandafter\ifx\csname natexlab\endcsname\relax\def\natexlab#1{#1}\fi

\bibitem[{{Adelman-McCarthy} {et~al.}(2008)}]{adelman08}
{Adelman-McCarthy}, J.~K. {et~al.} 2008, \apjs, 175, 297

\bibitem[{{Bergeron} {et~al.}(1995){Bergeron}, {Wesemael}, \&
  {Beauchamp}}]{bergeron95}
{Bergeron}, P., {Wesemael}, F., \& {Beauchamp}, A. 1995, \pasp, 107, 1047

\bibitem[{{Brown} {et~al.}(2009{\natexlab{a}}){Brown}, {Geller}, \&
  {Kenyon}}]{brown09a}
{Brown}, W.~R., {Geller}, M.~J., \& {Kenyon}, S.~J. 2009{\natexlab{a}}, \apj,
  690, 1639

\bibitem[{{Brown} {et~al.}(2009{\natexlab{b}}){Brown}, {Geller}, {Kenyon}, \&
  {Bromley}}]{brown09b}
{Brown}, W.~R., {Geller}, M.~J., {Kenyon}, S.~J., \& {Bromley}, B.~C.
  2009{\natexlab{b}}, \apjl, 690, L69

\bibitem[{{Brown} {et~al.}(2005){Brown}, {Geller}, {Kenyon}, \&
  {Kurtz}}]{brown05}
{Brown}, W.~R., {Geller}, M.~J., {Kenyon}, S.~J., \& {Kurtz}, M.~J. 2005,
  \apjl, 622, L33

\bibitem[{{Brown} {et~al.}(2006{\natexlab{a}}){Brown}, {Geller}, {Kenyon}, \&
  {Kurtz}}]{brown06}
---. 2006{\natexlab{a}}, \apjl, 640, L35

\bibitem[{{Brown} {et~al.}(2006{\natexlab{b}}){Brown}, {Geller}, {Kenyon}, \&
  {Kurtz}}]{brown06b}
---. 2006{\natexlab{b}}, \apj, 647, 303

\bibitem[{{Brown} {et~al.}(2007{\natexlab{a}}){Brown}, {Geller}, {Kenyon},
  {Kurtz}, \& {Bromley}}]{brown07a}
{Brown}, W.~R., {Geller}, M.~J., {Kenyon}, S.~J., {Kurtz}, M.~J., \& {Bromley},
  B.~C. 2007{\natexlab{a}}, \apj, 660, 311

\bibitem[{{Brown} {et~al.}(2007{\natexlab{b}}){Brown}, {Geller}, {Kenyon},
  {Kurtz}, \& {Bromley}}]{brown07b}
---. 2007{\natexlab{b}}, \apj, 671, 1708

\bibitem[{{Brown} {et~al.}(2010){Brown}, {Kilic}, {Allende Prieto}, \&
  {Kenyon}}]{brown10d}
{Brown}, W.~R., {Kilic}, M., {Allende Prieto}, C., \& {Kenyon}, S.~J. 2010,
  \mnras, submitted

\bibitem[{{Dong} {et~al.}(2010){Dong}, {Wang}, {Lin}, \& {Liu}}]{dong10}
{Dong}, R., {Wang}, Y., {Lin}, D.~N.~C., \& {Liu}, X. 2010, \apj, 715, 1036

\bibitem[{{Eisenstein} {et~al.}(2006)}]{eisenstein06}
{Eisenstein}, D.~J. {et~al.} 2006, \apjs, 167, 40

\bibitem[{{Fabricant} {et~al.}(1998){Fabricant}, {Cheimets}, {Caldwell}, \&
  {Geary}}]{fabricant98}
{Fabricant}, D., {Cheimets}, P., {Caldwell}, N., \& {Geary}, J. 1998, \pasp,
  110, 79

\bibitem[{{Girardi} {et~al.}(2004){Girardi}, {Grebel}, {Odenkirchen}, \&
  {Chiosi}}]{girardi04}
{Girardi}, L., {Grebel}, E.~K., {Odenkirchen}, M., \& {Chiosi}, C. 2004, \aap,
  422, 205

\bibitem[{{Girardi} {et~al.}(2002)}]{girardi02}
{Girardi}, L. {et~al.} 2002, \aap, 391, 195

\bibitem[{{Heber}(2009)}]{heber09}
{Heber}, U. 2009, \araa, 47, 211

\bibitem[{{Holberg} \& {Bergeron}(2006)}]{holberg06}
{Holberg}, J.~B. \& {Bergeron}, P. 2006, \aj, 132, 1221

\bibitem[{{Kawka} \& {Vennes}(2009)}]{kawka09}
{Kawka}, A. \& {Vennes}, S. 2009, \aap, 506, L25

\bibitem[{{Kawka} {et~al.}(2006){Kawka}, {Vennes}, {Oswalt}, {Smith}, \&
  {Silvestri}}]{kawka06}
{Kawka}, A., {Vennes}, S., {Oswalt}, T.~D., {Smith}, J.~A., \& {Silvestri},
  N.~M. 2006, \apjl, 643, L123

\bibitem[{{Kawka} {et~al.}(2010){Kawka}, {Vennes}, \& {Vaccaro}}]{kawka10}
{Kawka}, A., {Vennes}, S., \& {Vaccaro}, T.~R. 2010, \aap, 516, L7

\bibitem[{{Kenyon} \& {Garcia}(1986)}]{kenyon86}
{Kenyon}, S.~J. \& {Garcia}, M.~R. 1986, \aj, 91, 125

\bibitem[{{Kepler} {et~al.}(2007){Kepler}, {Kleinman}, {Nitta}, {Koester},
  {Castanheira}, {Giovannini}, {Costa}, \& {Althaus}}]{kepler07}
{Kepler}, S.~O., {Kleinman}, S.~J., {Nitta}, A., {Koester}, D., {Castanheira},
  B.~G., {Giovannini}, O., {Costa}, A.~F.~M., \& {Althaus}, L. 2007, \mnras,
  375, 1315

\bibitem[{{Kilic} {et~al.}(2010{\natexlab{a}}){Kilic}, {Allende Prieto},
  {Brown}, {Agueros}, {Kenyon}, \& {Camilo}}]{kilic10b}
{Kilic}, M., {Allende Prieto}, C., {Brown}, W.~R., {Agueros}, M.~A., {Kenyon},
  S.~J., \& {Camilo}, F. 2010{\natexlab{a}}, \apj, accepted

\bibitem[{{Kilic} {et~al.}(2007{\natexlab{a}}){Kilic}, {Allende Prieto},
  {Brown}, \& {Koester}}]{kilic07}
{Kilic}, M., {Allende Prieto}, C., {Brown}, W.~R., \& {Koester}, D.
  2007{\natexlab{a}}, \apj, 660, 1451

\bibitem[{{Kilic} {et~al.}(2010{\natexlab{b}}){Kilic}, {Brown}, {Allende
  Prieto}, {Kenyon}, \& {Panei}}]{kilic10}
{Kilic}, M., {Brown}, W.~R., {Allende Prieto}, C., {Kenyon}, S.~J., \& {Panei},
  J.~A. 2010{\natexlab{b}}, \apj, 716, 122

\bibitem[{{Kilic} {et~al.}(2007{\natexlab{b}}){Kilic}, {Brown}, {Allende
  Prieto}, {Pinsonneault}, \& {Kenyon}}]{kilic07b}
{Kilic}, M., {Brown}, W.~R., {Allende Prieto}, C., {Pinsonneault}, M., \&
  {Kenyon}, S. 2007{\natexlab{b}}, \apj, 664, 1088

\bibitem[{{Kilic} {et~al.}(2009){Kilic}, {Brown}, {Allende Prieto}, {Swift},
  {Kenyon}, {Liebert}, \& {Ag{\"u}eros}}]{kilic09}
{Kilic}, M., {Brown}, W.~R., {Allende Prieto}, C., {Swift}, B., {Kenyon},
  S.~J., {Liebert}, J., \& {Ag{\"u}eros}, M.~A. 2009, \apjl, 695, L92

\bibitem[{{Kilic} {et~al.}(2007{\natexlab{c}}){Kilic}, {Stanek}, \&
  {Pinsonneault}}]{kilic07c}
{Kilic}, M., {Stanek}, K.~Z., \& {Pinsonneault}, M.~H. 2007{\natexlab{c}},
  \apj, 671, 761

\bibitem[{{Koester}(2008)}]{koester08}
{Koester}, D. 2008, ArXiv:0812.0482

\bibitem[{{Koester} \& {Wilken}(2006)}]{koester06}
{Koester}, D. \& {Wilken}, D. 2006, \aap, 453, 1051

\bibitem[{{Kulkarni} \& {van Kerkwijk}(2010)}]{kulkarni10}
{Kulkarni}, S.~R. \& {van Kerkwijk}, M.~H. 2010, \apj, 719, 1123

\bibitem[{{Kurtz} \& {Mink}(1998)}]{kurtz98}
{Kurtz}, M.~J. \& {Mink}, D.~J. 1998, \pasp, 110, 934

\bibitem[{{Landau} \& {Lifshitz}(1958)}]{landau58}
{Landau}, L.~D. \& {Lifshitz}, E.~M. 1958, {The classical theory of fields}
  (Oxford: Pergamon Press)

\bibitem[{{Liebert} {et~al.}(2004){Liebert}, {Bergeron}, {Eisenstein},
  {Harris}, {Kleinman}, {Nitta}, \& {Krzesinski}}]{liebert04}
{Liebert}, J., {Bergeron}, P., {Eisenstein}, D., {Harris}, H.~C., {Kleinman},
  S.~J., {Nitta}, A., \& {Krzesinski}, J. 2004, \apjl, 606, L147

\bibitem[{{Marsh} {et~al.}(1995){Marsh}, {Dhillon}, \& {Duck}}]{marsh95}
{Marsh}, T.~R., {Dhillon}, V.~S., \& {Duck}, S.~R. 1995, \mnras, 275, 828

\bibitem[{{Marsh} {et~al.}(2010){Marsh}, {Gaensicke}, {Steeghs}, {Southworth},
  {Koester}, {Harris}, \& {Merry}}]{marsh10}
{Marsh}, T.~R., {Gaensicke}, B.~T., {Steeghs}, D., {Southworth}, J., {Koester},
  D., {Harris}, V., \& {Merry}, L. 2010, \apjl, submitted

\bibitem[{{Massey} {et~al.}(1988){Massey}, {Strobel}, {Barnes}, \&
  {Anderson}}]{massey88}
{Massey}, P., {Strobel}, K., {Barnes}, J.~V., \& {Anderson}, E. 1988, \apj,
  328, 315

\bibitem[{{Mullally} {et~al.}(2009){Mullally}, {Badenes}, {Thompson}, \&
  {Lupton}}]{mullally09}
{Mullally}, F., {Badenes}, C., {Thompson}, S.~E., \& {Lupton}, R. 2009, \apjl,
  707, L51

\bibitem[{{Munn} {et~al.}(2004)}]{munn04}
{Munn}, J.~A. {et~al.} 2004, \aj, 127, 3034

\bibitem[{{Panei} {et~al.}(2007){Panei}, {Althaus}, {Chen}, \& {Han}}]{panei07}
{Panei}, J.~A., {Althaus}, L.~G., {Chen}, X., \& {Han}, Z. 2007, \mnras, 382,
  779

\bibitem[{{Perets}(2010)}]{perets10b}
{Perets}, H.~B. 2010, \apj, submitted

\bibitem[{{Sch{\"o}nrich} {et~al.}(2010){Sch{\"o}nrich}, {Binney}, \&
  {Dehnen}}]{schonrich10}
{Sch{\"o}nrich}, R., {Binney}, J., \& {Dehnen}, W. 2010, \mnras, 403, 1829

\bibitem[{{Steinfadt} {et~al.}(2010){Steinfadt}, {Kaplan}, {Shporer},
  {Bildsten}, \& {Howell}}]{steinfadt10}
{Steinfadt}, J.~D.~R., {Kaplan}, D.~L., {Shporer}, A., {Bildsten}, L., \&
  {Howell}, S.~B. 2010, \apjl, 716, L146

\end{thebibliography}

\appendix \section{DATA TABLE}

	Table \ref{tab:dat} presents our radial velocity measurements. The Table
columns include object name, heliocentric Julian date, heliocentric radial velocity,
and velocity error.

\begin{deluxetable}{lcr}
\tabletypesize{\footnotesize}
\tablecolumns{3}
\tablewidth{0pt}
\tablecaption{Radial Velocity Measurements\label{tab:dat}}
\tablehead{
	\colhead{Object}& \colhead{HJD}& \colhead{$v_{helio}$}\\
			& +2450000     & (km s$^{-1}$)
}
	\startdata
J0755+4906 & 5150.867780 & $  393.5 \pm 17.5 $ \\
\nodata    & 5267.639886 & $  395.4 \pm 12.6 $ \\
\nodata    & 5267.654862 & $   27.8 \pm 11.5 $ \\
\nodata    & 5267.668518 & $ -421.8 \pm 18.0 $ \\
\nodata    & 5268.675038 & $ -363.5 \pm 23.8 $ \\
\nodata    & 5268.685061 & $ -490.9 \pm 19.5 $ \\
\nodata    & 5268.703821 & $  217.5 \pm 22.8 $ \\
\nodata    & 5268.713913 & $  384.5 \pm 22.0 $ \\
\nodata    & 5269.617361 & $ -358.8 \pm 22.0 $ \\
\nodata    & 5269.628715 & $ -386.8 \pm 31.4 $ \\
\nodata    & 5269.661999 & $  344.2 \pm 29.2 $ \\
\nodata    & 5269.674290 & $  -72.7 \pm 24.7 $ \\
\nodata    & 5276.680582 & $ -457.7 \pm 31.1 $ \\
\nodata    & 5276.693521 & $ -350.4 \pm 23.8 $ \\
\nodata    & 5276.704122 & $   93.0 \pm 18.9 $ \\
J0818+3536 & 5151.924119 & $  -92.7 \pm 16.2 $ \\
\nodata    & 5267.686783 & $  -52.8 \pm 13.9 $ \\
\nodata    & 5267.703298 & $  -55.3 \pm 17.7 $ \\
\nodata    & 5267.718598 & $  -15.1 \pm 14.4 $ \\
\nodata    & 5268.662172 & $ -225.0 \pm 15.4 $ \\
\nodata    & 5268.798134 & $  -28.8 \pm 12.4 $ \\
\nodata    & 5269.648138 & $ -350.8 \pm 19.7 $ \\
\nodata    & 5270.749626 & $ -273.1 \pm 17.5 $ \\
\nodata    & 5276.658219 & $ -107.1 \pm 30.6 $ \\
\nodata    & 5276.802073 & $ -290.2 \pm 20.7 $ \\
\nodata    & 5277.689765 & $ -287.2 \pm 31.7 $ \\
\nodata    & 5277.807903 & $ -155.8 \pm 24.1 $ \\
J0923+3028 & 3818.669260 & $  -22.3 \pm 13.4 $ \\
\nodata    & 5204.841296 & $  123.3 \pm 23.6 $ \\
\nodata    & 5204.849317 & $  364.1 \pm 29.1 $ \\
\nodata    & 5204.856366 & $  170.1 \pm 21.0 $ \\
\nodata    & 5204.864851 & $ -165.2 \pm 23.2 $ \\
\nodata    & 5204.871899 & $ -288.4 \pm 20.7 $ \\
\nodata    & 5204.880060 & $ -114.6 \pm 18.5 $ \\
\nodata    & 5204.887108 & $  101.8 \pm 24.9 $ \\
\nodata    & 5204.897757 & $  273.5 \pm 17.7 $ \\
\nodata    & 5204.904794 & $   15.5 \pm 18.6 $ \\
\nodata    & 5204.912596 & $ -198.1 \pm 16.4 $ \\
\nodata    & 5204.927423 & $  -27.7 \pm 16.8 $ \\
\nodata    & 5204.934460 & $  255.0 \pm 11.6 $ \\
\nodata    & 5204.942365 & $  245.2 \pm 21.3 $ \\
\nodata    & 5204.949403 & $   49.5 \pm 16.5 $ \\
\nodata    & 5204.957216 & $ -293.3 \pm 21.2 $ \\
\nodata    & 5204.964265 & $ -248.4 \pm 22.3 $ \\
\nodata    & 5204.972922 & $    2.3 \pm 20.3 $ \\
\nodata    & 5204.979960 & $  320.7 \pm 15.0 $ \\
\nodata    & 5204.987969 & $  260.3 \pm 21.7 $ \\
\nodata    & 5212.845177 & $  171.1 \pm 32.3 $ \\
\nodata    & 5212.858568 & $  183.6 \pm 45.1 $ \\
\nodata    & 5221.767105 & $ -248.2 \pm 28.5 $ \\
\nodata    & 5221.774790 & $ -307.1 \pm 19.1 $ \\
\nodata    & 5221.790045 & $  153.6 \pm 29.9 $ \\
\nodata    & 5221.797082 & $  316.9 \pm 35.1 $ \\
\nodata    & 5221.804744 & $   85.6 \pm 26.1 $ \\
\nodata    & 5221.811793 & $ -170.3 \pm 21.6 $ \\
\nodata    & 5221.819339 & $ -318.7 \pm 23.4 $ \\
\nodata    & 5221.826388 & $ -189.8 \pm 35.6 $ \\
\nodata    & 5221.834351 & $  144.9 \pm 17.4 $ \\
\nodata    & 5221.841400 & $  269.2 \pm 20.0 $ \\
\nodata    & 5221.848935 & $  133.3 \pm 13.0 $ \\
\nodata    & 5221.855983 & $ -142.1 \pm 15.5 $ \\
\nodata    & 5221.863657 & $ -282.4 \pm 12.6 $ \\
\nodata    & 5221.870706 & $ -153.8 \pm 18.7 $ \\
\nodata    & 5221.878275 & $  162.7 \pm 19.0 $ \\
\nodata    & 5221.885313 & $  300.7 \pm 17.0 $ \\
\nodata    & 5221.892859 & $  129.4 \pm 16.1 $ \\
\nodata    & 5221.899896 & $ -136.3 \pm 17.6 $ \\
\nodata    & 5221.907431 & $ -266.4 \pm 20.3 $ \\
\nodata    & 5221.914468 & $ -177.7 \pm 29.9 $ \\
\nodata    & 5221.922038 & $  -15.5 \pm 22.4 $ \\
\nodata    & 5221.929086 & $  285.6 \pm 18.0 $ \\
J1233+1602 & 5268.822025 & $ -351.5 \pm 10.6 $ \\
\nodata    & 5268.840648 & $ -157.5 \pm 13.1 $ \\
\nodata    & 5268.864607 & $  171.2 \pm  9.3 $ \\
\nodata    & 5268.887073 & $  301.6 \pm 15.5 $ \\
\nodata    & 5268.909573 & $  147.3 \pm  9.1 $ \\
\nodata    & 5269.793880 & $  290.8 \pm 14.6 $ \\
\nodata    & 5269.805316 & $  278.4 \pm 17.4 $ \\
\nodata    & 5269.816670 & $  151.1 \pm 15.8 $ \\
\nodata    & 5270.766544 & $ -344.9 \pm 14.1 $ \\
\nodata    & 5270.775294 & $ -347.4 \pm 20.0 $ \\
\nodata    & 5276.854668 & $   97.5 \pm 16.4 $ \\
\nodata    & 5276.866474 & $  191.1 \pm 12.1 $ \\
\nodata    & 5276.875895 & $  298.3 \pm 12.6 $ \\
J1422+4352 & 4596.887690 & $ -290.1 \pm 10.9 $ \\
\nodata    & 5268.969670 & $ -364.8 \pm 12.5 $ \\
\nodata    & 5268.978929 & $ -331.1 \pm  9.8 $ \\
\nodata    & 5268.988397 & $ -352.3 \pm 13.6 $ \\
\nodata    & 5270.854102 & $ -373.6 \pm 14.8 $ \\
\nodata    & 5384.771754 & $  -93.0 \pm 15.4 $ \\
\nodata    & 5384.793095 & $  -36.8 \pm 15.6 $ \\
\nodata    & 5385.726490 & $ -314.3 \pm 21.4 $ \\
J1439+1002 & 3882.885571 & $   -8.6 \pm  7.9 $ \\
\nodata    & 4922.892856 & $  -21.0 \pm  6.9 $ \\
\nodata    & 4922.892856 & $   82.0 \pm  4.5 $ \\
\nodata    & 4922.977535 & $  116.4 \pm  5.6 $ \\
\nodata    & 4923.020372 & $  137.9 \pm 18.8 $ \\
\nodata    & 4923.973820 & $ -126.2 \pm  7.4 $ \\
\nodata    & 4924.020430 & $ -191.7 \pm 18.9 $ \\
\nodata    & 5267.943738 & $  -56.8 \pm 14.4 $ \\
\nodata    & 5267.969654 & $  -50.0 \pm 11.2 $ \\
\nodata    & 5268.036672 & $  -15.3 \pm 25.6 $ \\
\nodata    & 5268.957759 & $  111.4 \pm  7.0 $ \\
\nodata    & 5268.962158 & $  110.6 \pm  5.1 $ \\
\nodata    & 5269.971779 & $ -147.0 \pm 14.4 $ \\
\nodata    & 5269.977266 & $ -165.6 \pm 11.1 $ \\
\nodata    & 5270.959442 & $ -189.1 \pm  5.7 $ \\
\nodata    & 5270.979860 & $ -178.5 \pm  9.1 $ \\
\nodata    & 5271.011239 & $ -119.8 \pm  9.1 $ \\
\nodata    & 5275.934512 & $  120.3 \pm 10.2 $ \\
\nodata    & 5277.886834 & $ -216.8 \pm 10.5 $ \\
\nodata    & 5384.660342 & $ -246.8 \pm 10.3 $ \\
\nodata    & 5385.755771 & $  202.7 \pm 47.3 $ \\
J1448+1342 & 4237.919202 & $  -21.6 \pm 15.6 $ \\
\nodata    & 5276.885957 & $  -38.5 \pm 19.0 $ \\
\nodata    & 5276.892265 & $   -0.1 \pm 25.0 $ \\
\nodata    & 5276.900298 & $  -22.9 \pm 17.6 $ \\
\nodata    & 5276.906641 & $  -39.8 \pm 23.6 $ \\
\nodata    & 5277.023591 & $ -182.0 \pm 69.6 $ \\
\nodata    & 5277.896227 & $   -3.2 \pm 25.8 $ \\
\nodata    & 5277.902686 & $  -14.6 \pm 27.1 $ \\
\nodata    & 5277.911552 & $  -49.7 \pm 32.8 $ \\
\nodata    & 5277.917988 & $    4.5 \pm 20.8 $ \\
\nodata    & 5277.990376 & $  -53.4 \pm 22.3 $ \\
\nodata    & 5277.996811 & $  -82.0 \pm 27.8 $ \\
\nodata    & 5278.004983 & $  -35.8 \pm 20.4 $ \\
\nodata    & 5278.011256 & $  -23.8 \pm 16.6 $ \\
\nodata    & 5384.666893 & $   -0.3 \pm 16.7 $ \\
\nodata    & 5384.752430 & $  -56.4 \pm 20.0 $ \\
\nodata    & 5385.762534 & $  -12.3 \pm 32.9 $ \\
J1512+2615 & 3879.873052 & $ -127.7 \pm 11.2 $ \\
\nodata    & 5267.949987 & $   56.7 \pm 17.4 $ \\
\nodata    & 5267.974827 & $   87.6 \pm 19.0 $ \\
\nodata    & 5267.997247 & $   49.5 \pm 18.6 $ \\
\nodata    & 5269.017902 & $  -96.7 \pm 14.4 $ \\
\nodata    & 5269.023840 & $ -101.2 \pm 17.1 $ \\
\nodata    & 5270.006995 & $  -74.1 \pm 19.5 $ \\
\nodata    & 5271.002158 & $   66.2 \pm 16.6 $ \\
\nodata    & 5271.017749 & $    1.0 \pm 28.4 $ \\
\nodata    & 5271.024775 & $   53.1 \pm 26.8 $ \\
\nodata    & 5276.915247 & $   42.9 \pm 17.9 $ \\
\nodata    & 5277.013574 & $  129.0 \pm 31.5 $ \\
\nodata    & 5277.943822 & $ -154.4 \pm 25.1 $ \\
\nodata    & 5384.675640 & $ -189.8 \pm 22.7 $ \\
\nodata    & 5384.763249 & $ -127.7 \pm 15.2 $ \\
\nodata    & 5385.747861 & $   22.7 \pm 28.1 $ \\
J1630+2712 & 5009.892692 & $ -239.7 \pm 18.9 $ \\
\nodata    & 5267.958434 & $  -47.1 \pm 18.0 $ \\
\nodata    & 5267.983829 & $   24.0 \pm 14.3 $ \\
\nodata    & 5268.011944 & $  103.0 \pm 14.5 $ \\
\nodata    & 5269.005570 & $ -280.4 \pm 13.7 $ \\
\nodata    & 5269.985526 & $   17.1 \pm 15.1 $ \\
\nodata    & 5270.988168 & $  -80.9 \pm 14.5 $ \\
\nodata    & 5276.925057 & $ -133.0 \pm 19.3 $ \\
\nodata    & 5276.999529 & $ -332.8 \pm 29.4 $ \\
\nodata    & 5384.809434 & $ -345.1 \pm 15.2 $ \\
\nodata    & 5384.868187 & $ -197.8 \pm 31.1 $ \\
\nodata    & 5385.781067 & $  126.8 \pm 13.6 $ \\
\nodata    & 5385.842754 & $  -58.9 \pm 34.1 $ \\
J2119$-$0018 & 5008.884992 & $  -15.2 \pm 18.7 $ \\
\nodata    & 5152.578978 & $    7.8 \pm 10.5 $ \\
\nodata    & 5152.591870 & $  266.2 \pm 10.8 $ \\
\nodata    & 5153.581786 & $ -203.5 \pm 16.9 $ \\
\nodata    & 5153.590153 & $ -319.1 \pm 21.7 $ \\
\nodata    & 5153.597282 & $ -396.7 \pm 15.3 $ \\
\nodata    & 5153.605372 & $ -329.3 \pm 18.1 $ \\
\nodata    & 5153.612500 & $ -159.7 \pm 15.1 $ \\
\nodata    & 5153.628726 & $  306.2 \pm 23.9 $ \\
\nodata    & 5153.636966 & $  357.4 \pm 17.9 $ \\
\nodata    & 5153.645148 & $  312.6 \pm 18.3 $ \\
\nodata    & 5153.654661 & $  154.2 \pm 28.5 $ \\
\nodata    & 5153.663884 & $ -115.9 \pm 29.2 $ \\
\nodata    & 5157.626210 & $  383.6 \pm 31.6 $ \\
\nodata    & 5157.635329 & $  343.2 \pm 16.8 $ \\
\nodata    & 5157.644287 & $  265.7 \pm 17.8 $ \\
\nodata    & 5157.653291 & $   26.0 \pm 28.6 $ \\
\nodata    & 5157.662260 & $ -139.8 \pm 25.8 $ \\
\nodata    & 5157.672131 & $ -427.6 \pm 17.2 $ \\
\nodata    & 5158.586060 & $  337.8 \pm 21.8 $ \\
\nodata    & 5158.594878 & $  312.5 \pm 16.8 $ \\
\nodata    & 5158.603454 & $  150.1 \pm 30.5 $ \\
\nodata    & 5158.613036 & $ -173.7 \pm 25.2 $ \\
\nodata    & 5158.625477 & $ -383.8 \pm 17.5 $ \\
\nodata    & 5158.636229 & $ -443.9 \pm 18.3 $ \\
	\enddata
\end{deluxetable}

\end{document}